\PassOptionsToPackage{final}{hyperref}
\documentclass[letterpaper,sigconf]{acmart}
\AtBeginDocument{%
  }

\usepackage[acronyms,nonumberlist,nopostdot,nomain,nogroupskip,acronymlists={hidden}]{glossaries}
\newglossary[algh]{hidden}{acrh}{acnh}{Hidden Acronyms}
\newacronym{3gpp}{3GPP}{3rd Generation Partnership Project}
\newacronym{4g}{4G}{4th generation}
\newacronym{5g}{5G}{5th generation}
\newacronym{6g}{6G}{6th generation}
\newacronym{5gc}{5GC}{5G Core}
\newacronym{adc}{ADC}{Analog to Digital Converter}
\newacronym{aerpaw}{AERPAW}{Aerial Experimentation and Research Platform for Advanced Wireless}
\newacronym{ai}{AI}{Artificial Intelligence}
\newacronym{aimd}{AIMD}{Additive Increase Multiplicative Decrease}
\newacronym{am}{AM}{Acknowledged Mode}
\newacronym{amc}{AMC}{Adaptive Modulation and Coding}
\newacronym{amf}{AMF}{Access and Mobility Management Function}
\newacronym{aops}{AOPS}{Adaptive Order Prediction Scheduling}
\newacronym{api}{API}{Application Programming Interface}
\newacronym{apn}{APN}{Access Point Name}
\newacronym{ap}{AP}{Application Protocol}
\newacronym{aqm}{AQM}{Active Queue Management}
\newacronym{ausf}{AUSF}{Authentication Server Function}
\newacronym{avc}{AVC}{Advanced Video Coding}
\newacronym{awgn}{AGWN}{Additive White Gaussian Noise}
\newacronym{balia}{BALIA}{Balanced Link Adaptation Algorithm}
\newacronym{bbu}{BBU}{Base Band Unit}
\newacronym{bdp}{BDP}{Bandwidth-Delay Product}
\newacronym{ber}{BER}{Bit Error Rate}
\newacronym{bf}{BF}{Beamforming}
\newacronym{bler}{BLER}{Block Error Rate}
\newacronym{brr}{BRR}{Bayesian Ridge Regressor}
\newacronym{bs}{BS}{Base Station}
\newacronym{bsr}{BSR}{Buffer Status Report}
\newacronym{bss}{BSS}{Business Support System}
\newacronym{bts}{BTS}{Base Transceiver Station}
\newacronym{ca}{CA}{Carrier Aggregation}
\newacronym{caas}{CaaS}{Connectivity-as-a-Service}
\newacronym{cb}{CB}{Code Block}
\newacronym{cbr}{CBR}{Collateral Blocking Rate}
\newacronym{cc}{CC}{Congestion Control}
\newacronym{ccid}{CCID}{Congestion Control ID}
\newacronym{cco}{CC}{Carrier Component}
\newacronym{cdd}{CDD}{Cyclic Delay Diversity}
\newacronym{cdf}{CDF}{Cumulative Distribution Function}
\newacronym{cdn}{CDN}{Content Distribution Network}
\newacronym{cia}{CIA}{Confidentiality-Integrity-Availability}
\newacronym{cir}{CIR}{Channel Impulse Response}
\newacronym{cli}{CLI}{Command-line Interface}
\newacronym{cn}{CN}{Core Network}
\newacronym{cnn}{CNN}{Convolutional Neural Network}
\newacronym{codel}{CoDel}{Controlled Delay Management}
\newacronym{comac}{COMAC}{Converged Multi-Access and Core}
\newacronym{cord}{CORD}{Central Office Re-architected as a Datacenter}
\newacronym{cornet}{CORNET}{COgnitive Radio NETwork}
\newacronym{cosmos}{COSMOS}{Cloud Enhanced Open Software Defined Mobile Wireless Testbed for City-Scale Deployment}
\newacronym{cots}{COTS}{Commercial Off-the-Shelf}
\newacronym{cp}{CP}{Control Plane}
\newacronym{cyp}{CP}{Cyclic Prefix}
\newacronym{up}{UP}{User Plane}
\newacronym{cpu}{CPU}{Central Processing Unit}
\newacronym{cqi}{CQI}{Channel Quality Information}
\newacronym{cql}{CQL}{Conservative Q-Learning}
\newacronym{cr}{CR}{Cognitive Radio}
\newacronym{cran}{CRAN}{Cloud \gls{ran}}
\newacronym{crs}{CRS}{Cell Reference Signal}
\newacronym{csi}{CSI}{Channel State Information}
\newacronym{csirs}{CSI-RS}{Channel State Information - Reference Signal}
\newacronym{csv}{CSV}{Comma Separated Values}
\newacronym{cu}{CU}{Central Unit}
\newacronym{cucp}{CU-CP}{Central Unit Control Plane}
\newacronym{cuup}{CU-UP}{Central Unit User Plane}
\newacronym{d2tcp}{D$^2$TCP}{Deadline-aware Data center TCP}
\newacronym{d3}{D$^3$}{Deadline-Driven Delivery}
\newacronym{dac}{DAC}{Digital to Analog Converter}
\newacronym{dag}{DAG}{Directed Acyclic Graph}
\newacronym{dapp}{DApp}{Distributed Application}
\newacronym{dapps}{DApps}{Distributed Applications}
\newacronym{das}{DAS}{Distributed Antenna System}
\newacronym{dash}{DASH}{Dynamic Adaptive Streaming over HTTP}
\newacronym{dbscan}{DBSCAN}{Density-Based Spatial
Clustering of Applications with Noise}
\newacronym{dc}{DC}{Dual Connectivity}
\newacronym{dccp}{DCCP}{Datagram Congestion Control Protocol}
\newacronym{dce}{DCE}{Direct Code Execution}
\newacronym{dci}{DCI}{Downlink Control Information}
\newacronym{dctcp}{DCTCP}{Data Center TCP}
\newacronym{dl}{DL}{Downlink}
\newacronym{dmr}{DMR}{Deadline Miss Ratio}
\newacronym{dmrs}{DMRS}{DeModulation Reference Signal}
\newacronym{dos}{DoS}{Denial-of-Service}
\newacronym{dqn}{DQN}{Deep Q-Network}
\newacronym{dtw}{DTW}{Dynamic Time Warping}
\newacronym{drlcc}{DRL-CC}{Deep Reinforcement Learning Congestion Control}
\newacronym{drs}{DRS}{Discovery Reference Signal}
\newacronym{du}{DU}{Distributed Unit}
\newacronym{ee}{EE}{Energy Efficiency}
\newacronym{e2e}{E2E}{end-to-end}
\newacronym{e2sm}{E2SM}{E2 Service Model}
\newacronym{earfcn}{EARFCN}{E-UTRA Absolute Radio Frequency Channel Number}
\newacronym{ecaas}{ECaaS}{Edge-Cloud-as-a-Service}
\newacronym{ecn}{ECN}{Explicit Congestion Notification}
\newacronym{edf}{EDF}{Earliest Deadline First}
\newacronym{embb}{eMBB}{Enhanced Mobile Broadband}
\newacronym{empower}{EMPOWER}{EMpowering transatlantic PlatfOrms for advanced WirEless Research}
\newacronym{enb}{eNB}{evolved Node Base}
\newacronym{endc}{EN-DC}{E-UTRAN-\gls{nr} \gls{dc}}
\newacronym{epc}{EPC}{Evolved Packet Core}
\newacronym{eps}{EPS}{Evolved Packet System}
\newacronym{es}{ES}{Edge Server}
\newacronym{etsi}{ETSI}{European Telecommunications Standards Institute}
\newacronym[firstplural=Estimated Times of Arrival (ETAs)]{eta}{ETA}{Estimated Time of Arrival}
\newacronym{eutran}{E-UTRAN}{Evolved Universal Terrestrial Access Network}
\newacronym{faas}{FaaS}{Function-as-a-Service}
\newacronym{fapi}{FAPI}{Functional Application Platform Interface}
\newacronym{fdd}{FDD}{Frequency Division Duplexing}
\newacronym{fdm}{FDM}{Frequency Division Multiplexing}
\newacronym{fdma}{FDMA}{Frequency Division Multiple Access}
\newacronym{fed4fire}{FED4FIRE+}{Federation 4 Future Internet Research and Experimentation Plus}
\newacronym{fir}{FIR}{Finite Impulse Response}
\newacronym{fit}{FIT}{Future \acrlong{iot}}
\newacronym{fpga}{FPGA}{Field Programmable Gate Array}
\newacronym{fr1}{FR1}{Frequency Range 1}
\newacronym{fr2}{FR2}{Frequency Range 2}
\newacronym{fs}{FS}{Fast Switching}
\newacronym{fscc}{FSCC}{Flow Sharing Congestion Control}
\newacronym{ftp}{FTP}{File Transfer Protocol}
\newacronym{fw}{FW}{Flow Window}
\newacronym{ge}{GE}{Gaussian Elimination}
\newacronym{gnb}{gNB}{Next Generation Node Base}
\newacronym{gop}{GOP}{Group of Pictures}
\newacronym{gpr}{GPR}{Gaussian Process Regressor}
\newacronym{gpu}{GPU}{Graphics Processing Unit}
\newacronym{gtp}{GTP}{GPRS Tunneling Protocol}
\newacronym{gan}{GAN}{Generative Adversarial Network}
\newacronym{gui}{GUI}{Graphical User Interface}
\newacronym{gtpc}{GTP-C}{GPRS Tunnelling Protocol Control Plane}
\newacronym{gtpu}{GTP-U}{GPRS Tunnelling Protocol User Plane}
\newacronym{gtpv2c}{GTPv2-C}{\gls{gtp} v2 - Control}
\newacronym{gw}{GW}{Gateway}
\newacronym{harq}{HARQ}{Hybrid Automatic Repeat reQuest}
\newacronym{hetnet}{HetNet}{Heterogeneous Network}
\newacronym{hh}{HH}{Hard Handover}
\newacronym{hol}{HOL}{Head-of-Line}
\newacronym{hqf}{HQF}{Highest-quality-first}
\newacronym{hss}{HSS}{Home Subscription Server}
\newacronym{http}{HTTP}{HyperText Transfer Protocol}
\newacronym{ia}{IA}{Initial Access}
\newacronym{iab}{IAB}{Integrated Access and Backhaul}
\newacronym{ic}{IC}{Incident Command}
\newacronym{ietf}{IETF}{Internet Engineering Task Force}
\newacronym{imsi}{IMSI}{International Mobile Subscriber Identity}
\newacronym{imt}{IMT}{International Mobile Telecommunication}
\newacronym{iot}{IoT}{Internet of Things}
\newacronym{ip}{IP}{Internet Protocol}
\newacronym{ipc}{IPC}{inter-process communication}
\newacronym{itu}{ITU}{International Telecommunication Union}
\newacronym{knn}{KNN}{k-nearest neighbors}
\newacronym{kpi}{KPI}{Key Performance Indicator}
\newacronym{kpm}{KPM}{Key Performance Metric}
\newacronym{kvm}{KVM}{Kernel-based Virtual Machine}
\newacronym{llm}{LLM}{Large Language Model}
\newacronym{los}{LOS}{Line-of-Sight}
\newacronym{lsm}{LSM}{Link-to-System Mapping}
\newacronym{lstm}{LSTM}{Long Short Term Memory}
\newacronym{lte}{LTE}{Long Term Evolution}
\newacronym{lxc}{LXC}{Linux Container}
\newacronym{m2m}{M2M}{Machine to Machine}
\newacronym{mac}{MAC}{Medium Access Control}
\newacronym{mue}{MUE}{Malicious UE}
\newacronym{manet}{MANET}{Mobile Ad Hoc Network}
\newacronym{mano}{MANO}{Management and Orchestration}
\newacronym{mc}{MC}{Multi-Connectivity}
\newacronym{mcc}{MCC}{Mobile Cloud Computing}
\newacronym{mchem}{MCHEM}{Massive Channel Emulator}
\newacronym{mcs}{MCS}{Modulation and Coding Scheme}
\newacronym{mec}{MEC}{Multi-access Edge Computing}
\newacronym{mec2}{MEC}{Mobile Edge Cloud}
\newacronym{mfc}{MFC}{Mobile Fog Computing}
\newacronym{mgen}{MGEN}{Multi-Generator}
\newacronym{mi}{MI}{Mutual Information}
\newacronym{mib}{MIB}{Master Information Block}
\newacronym{miesm}{MIESM}{Mutual Information Based Effective SINR}
\newacronym{mimo}{MIMO}{Multiple Input, Multiple Output}
\newacronym{ml}{ML}{Machine Learning}
\newacronym{mlr}{MLR}{Maximum-local-rate}
\newacronym[plural=\gls{mme}s,firstplural=Mobility Management Entities (MMEs)]{mme}{MME}{Mobility Management Entity}
\newacronym{mmtc}{mMTC}{Massive Machine-Type Communications}
\newacronym{mmwave}{mmWave}{millimeter wave}
\newacronym{mpdccp}{MP-DCCP}{Multipath Datagram Congestion Control Protocol}
\newacronym{mptcp}{MPTCP}{Multipath TCP}
\newacronym{mr}{MR}{Maximum Rate}
\newacronym{mrdc}{MR-DC}{Multi \gls{rat} \gls{dc}}
\newacronym{mse}{MSE}{Mean Square Error}
\newacronym{mss}{MSS}{Maximum Segment Size}
\newacronym{mt}{MT}{Mobile Termination}
\newacronym{mtd}{MTD}{Machine-Type Device}
\newacronym{mtu}{MTU}{Maximum Transmission Unit}
\newacronym{mumimo}{MU-MIMO}{Multi-user \gls{mimo}}
\newacronym{mvno}{MVNO}{Mobile Virtual Network Operator}
\newacronym{nalu}{NALU}{Network Abstraction Layer Unit}
\newacronym{nas}{NAS}{Non-Access Stratum}
\newacronym{nat}{NAT}{Network Address Translation}
\newacronym{nbiot}{NB-IoT}{Narrow Band IoT}
\newacronym{nearrt}{near-RT}{Near-Real-Time}
\newacronym{nic}{NIC}{Network Interface Card}
\newacronym{nonrt}{non-RT}{Non-Real-time}
\newacronym{nfv}{NFV}{Network Function Virtualization}
\newacronym{nfvi}{NFVI}{Network Function Virtualization Infrastructure}
\newacronym{ni}{NI}{Network Interfaces}
\newacronym{nlos}{NLOS}{Non-Line-of-Sight}
\newacronym{now}{NOW}{Non Overlapping Window}
\newacronym{nsm}{NSM}{Network Service Mesh}
\newacronym[type=hidden]{nr}{NR}{New Radio}
\newacronym{nextg}{NextG}{Next Generation}
\newacronym{nrf}{NRF}{Network Repository Function}
\newacronym{nsa}{NSA}{Non Stand Alone}
\newacronym{nse}{NSE}{Network Slicing Engine}
\newacronym{nssf}{NSSF}{Network Slice Selection Function}
\newacronym{ngrg}{nGRG}{next Generation Research Group}
\newacronym{o2i}{O2I}{Outdoor to Indoor}
\newacronym{oai}{OAI}{OpenAirInterface}
\newacronym{oaicn}{OAI-CN}{\gls{oai} \acrlong{cn}}
\newacronym{oairan}{OAI-RAN}{\acrlong{oai} \acrlong{ran}}
\newacronym{oam}{OAM}{Operations, Administration and Maintenance}
\newacronym{ofdm}{OFDM}{Orthogonal Frequency Division Multiplexing}
\newacronym{olia}{OLIA}{Opportunistic Linked Increase Algorithm}
\newacronym{omec}{OMEC}{Open Mobile Evolved Core}
\newacronym{onap}{ONAP}{Open Network Automation Platform}
\newacronym{onf}{ONF}{Open Networking Foundation}
\newacronym{onos}{ONOS}{Open Networking Operating System}
\newacronym{oom}{OOM}{\gls{onap} Operations Manager}
\newacronym{opnfv}{OPNFV}{Open Platform for \gls{nfv}}
\newacronym{orbit}{ORBIT}{Open-Access Research Testbed for Next-Generation Wireless Networks}
\newacronym{os}{OS}{Operating System}
\newacronym{osm2}{OSM}{Open Street Map}
\newacronym{oss}{OSS}{Operations Support System}
\newacronym{ota}{OTA}{Over-The-Air}
\newacronym{pa}{PA}{Position-aware}
\newacronym{pase}{PASE}{Prioritization, Arbitration, and Self-adjusting Endpoints}
\newacronym{pawr}{PAWR}{Platforms for Advanced Wireless Research}
\newacronym{pbch}{PBCH}{Physical Broadcast Channel}
\newacronym{pca}{PCA}{Principal Component Analysis}
\newacronym{psd}{PSD}{Power Spectral Density}
\newacronym{pps}{PPS}{Pulse-Per-Second}
\newacronym{pcef}{PCEF}{Policy and Charging Enforcement Function}
\newacronym{pcfich}{PCFICH}{Physical Control Format Indicator Channel}
\newacronym{pcrf}{PCRF}{Policy and Charging Rules Function}
\newacronym{pdcch}{PDCCH}{Physical Downlink Control Channel}
\newacronym{pdcp}{PDCP}{Packet Data Convergence Protocol}
\newacronym{pdsch}{PDSCH}{Physical Downlink Shared Channel}
\newacronym{pdu}{PDU}{Packet Data Unit}
\newacronym{pf}{PF}{Proportional Fair}
\newacronym{pgw}{PGW}{Packet Gateway}
\newacronym{phich}{PHICH}{Physical Hybrid ARQ Indicator Channel}
\newacronym{phy}{PHY}{Physical}
\newacronym{pl}{PL}{Path Loss}
\newacronym{pmch}{PMCH}{Physical Multicast Channel}
\newacronym{pmi}{PMI}{Precoding Matrix Indicators}
\newacronym{powder}{POWDER}{Platform for Open Wireless Data-driven Experimental Research}
\newacronym{ppo}{PPO}{Proximal Policy Optimization}
\newacronym{ppp}{PPP}{Poisson Point Process}
\newacronym{prach}{PRACH}{Physical Random Access Channel}
\newacronym{prb}{PRB}{Physical Resource Block}
\newacronym{psnr}{PSNR}{Peak Signal to Noise Ratio}
\newacronym{pss}{PSS}{Primary Synchronization Signal}
\newacronym{pucch}{PUCCH}{Physical Uplink Control Channel}
\newacronym{pusch}{PUSCH}{Physical Uplink Shared Channel}
\newacronym{qam}{QAM}{Quadrature Amplitude Modulation}
\newacronym{qci}{QCI}{\gls{qos} Class Identifier}
\newacronym{qoe}{QoE}{Quality of Experience}
\newacronym{qos}{QoS}{Quality of Service}
\newacronym{quic}{QUIC}{Quick UDP Internet Connections}
\newacronym{rach}{RACH}{Random Access Channel}
\newacronym{ram}{RAM}{Random Access Memory}
\newacronym{ran}{RAN}{Radio Access Network}
\newacronym[firstplural=Radio Access Technologies (RATs)]{rat}{RAT}{Radio Access Technology}
\newacronym{rbg}{RBG}{Resource Block Group}
\newacronym{rcn}{RCN}{Research Coordination Network}
\newacronym{rc}{RC}{RAN Control}
\newacronym{rec}{REC}{Radio Edge Cloud}
\newacronym{red}{RED}{Random Early Detection}
\newacronym{renew}{RENEW}{Reconfigurable Eco-system for Next-generation End-to-end Wireless}
\newacronym{rf}{RF}{Radio Frequency}
\newacronym{rfi}{RFI}{Radio Frequency Interference}
\newacronym{rfc}{RFC}{Request for Comments}
\newacronym{rfr}{RFR}{Random Forest Regressor}
\newacronym{ric}{RIC}{RAN Intelligent Controller}
\newacronym{rlc}{RLC}{Radio Link Control}
\newacronym{rlf}{RLF}{Radio Link Failure}
\newacronym{rlnc}{RLNC}{Random Linear Network Coding}
\newacronym{rmr}{RMR}{RIC Message Router}
\newacronym{rmse}{RMSE}{Root Mean Squared Error}
\newacronym{rnis}{RNIS}{Radio Network Information Service}
\newacronym{rnti}{RNTI}{Radio Network Temporary Identifier}
\newacronym{rr}{RR}{Round Robin}
\newacronym{rrc}{RRC}{Radio Resource Control}
\newacronym{rrm}{RRM}{Radio Resource Management}
\newacronym{rru}{RRU}{Remote Radio Unit}
\newacronym{rs}{RS}{Remote Server}
\newacronym{rsrp}{RSRP}{Reference Signal Received Power}
\newacronym{rsrq}{RSRQ}{Reference Signal Received Quality}
\newacronym{rss}{RSS}{Received Signal Strength}
\newacronym{rssi}{RSSI}{Received Signal Strength Indicator}
\newacronym{rt}{RT}{Real-time}
\newacronym{rtt}{RTT}{Round Trip Time}
\newacronym{ru}{RU}{Radio Unit}
\newacronym{rw}{RW}{Receive Window}
\newacronym{rx}{RX}{Receiver}
\newacronym{s1ap}{S1AP}{S1 Application Protocol}
\newacronym{sa}{SA}{standalone}
\newacronym{sack}{SACK}{Selective Acknowledgment}
\newacronym{sap}{SAP}{Service Access Point}
\newacronym{sc2}{SC2}{Spectrum Collaboration Challenge}
\newacronym{scef}{SCEF}{Service Capability Exposure Function}
\newacronym{sch}{SCH}{Secondary Cell Handover}
\newacronym{scoot}{SCOOT}{Split Cycle Offset Optimization Technique}
\newacronym{sctp}{SCTP}{Stream Control Transmission Protocol}
\newacronym{sdap}{SDAP}{Service Data Adaptation Protocol}
\newacronym{sdk}{SDK}{Software Development Kit}
\newacronym{sdm}{SDM}{Space Division Multiplexing}
\newacronym{sdma}{SDMA}{Spatial Division Multiple Access}
\newacronym{sdn}{SDN}{Software-defined Networking}
\newacronym{sdr}{SDR}{Software-defined Radio}
\newacronym{seba}{SEBA}{SDN-Enabled Broadband Access}
\newacronym{sgsn}{SGSN}{Serving GPRS Support Node}
\newacronym{sgw}{SGW}{Service Gateway}
\newacronym{si}{SI}{Study Item}
\newacronym{sib}{SIB}{Secondary Information Block}
\newacronym{sinr}{SINR}{Signal to Interference plus Noise Ratio}
\newacronym{sip}{SIP}{Session Initiation Protocol}
\newacronym{siso}{SISO}{Single Input, Single Output}
\newacronym{sla}{SLA}{Service Level Agreement}
\newacronym{sm}{SM}{Service Model}
\newacronym{smf}{SMF}{Session Management Function}
\newacronym{smo}{SMO}{Service Management and Orchestration}
\newacronym{sms}{SMS}{Short Message Service}
\newacronym{smsgmsc}{SMS-GMSC}{\gls{sms}-Gateway}
\newacronym{snr}{SNR}{Signal-to-Noise-Ratio}
\newacronym{son}{SON}{Self-Organizing Network}
\newacronym{sptcp}{SPTCP}{Single Path TCP}
\newacronym{srb}{SRB}{Service Radio Bearer}
\newacronym{swig}{SWIG}{Simplified Wrapper and Interface Generator}
\newacronym{srn}{SRN}{Standard Radio Node}
\newacronym{srs}{SRS}{Sounding Reference Signal}
\newacronym{ss}{SS}{Synchronization Signal}
\newacronym{sss}{SSS}{Secondary Synchronization Signal}
\newacronym{st}{ST}{Spanning Tree}
\newacronym{suci}{SUCI}{Subscription Concealed Identifier}
\newacronym{svc}{SVC}{Scalable Video Coding}
\newacronym{s-tmsi}{S-TMSI}{SAE Temporary Mobile Subscriber Identity}
\newacronym{ta}{TA}{Timing Advance}
\newacronym{tb}{TB}{Transport Block}
\newacronym{tcp}{TCP}{Transmission Control Protocol}
\newacronym{tdd}{TDD}{Time Division Duplexing}
\newacronym{tdl}{TDL}{Tapped Delay Line}
\newacronym{tdm}{TDM}{Time Division Multiplexing}
\newacronym{tdma}{TDMA}{Time Division Multiple Access}
\newacronym{tfl}{TfL}{Transport for London}
\newacronym{tfrc}{TFRC}{TCP-Friendly Rate Control}
\newacronym{tft}{TFT}{Traffic Flow Template}
\newacronym{tgen}{TGEN}{Traffic Generator}
\newacronym{tip}{TIP}{Telecom Infra Project}
\newacronym{tm}{TM}{Transparent Mode}
\newacronym{to}{TO}{Telco Operator}
\newacronym{tr}{TR}{Technical Report}
\newacronym{trp}{TRP}{Transmitter Receiver Pair}
\newacronym{ts}{TS}{Technical Specification}
\newacronym{tti}{TTI}{Transmission Time Interval}
\newacronym{ttt}{TTT}{Time-to-Trigger}
\newacronym{tx}{TX}{Transmitter}
\newacronym{uas}{UAS}{Unmanned Aerial System}
\newacronym{uav}{UAV}{Unmanned Aerial Vehicle}
\newacronym{udm}{UDM}{Unified Data Management}
\newacronym{udp}{UDP}{User Datagram Protocol}
\newacronym{udr}{UDR}{Unified Data Repository}
\newacronym{ue}{UE}{User Equipment}
\newacronym{uhd}{UHD}{\gls{usrp} Hardware Driver}
\newacronym{ul}{UL}{Uplink}
\newacronym{um}{UM}{Unacknowledged Mode}
\newacronym{umi}{UMi}{Urban Micro}
\newacronym{uml}{UML}{Unified Modeling Language}
\newacronym{upa}{UPA}{Uniform Planar Array}
\newacronym{upf}{UPF}{User Plane Function}
\newacronym{urllc}{URLLC}{Ultra Reliable and Low Latency Communications}
\newacronym{usa}{U.S.}{United States}
\newacronym{usim}{USIM}{Universal Subscriber Identity Module}
\newacronym{usrp}{USRP}{Universal Software Radio Peripheral}
\newacronym{utc}{UTC}{Urban Traffic Control}
\newacronym{vim}{VIM}{Virtualization Infrastructure Manager}
\newacronym{vm}{VM}{Virtual Machine}
\newacronym{vnf}{VNF}{Virtual Network Function}
\newacronym{volte}{VoLTE}{Voice over \gls{lte}}
\newacronym{voltha}{VOLTHA}{Virtual OLT HArdware Abstraction}
\newacronym{vr}{VR}{Virtual Reality}
\newacronym{vran}{vRAN}{Virtualized \gls{ran}}
\newacronym{vss}{VSS}{Video Streaming Server}
\newacronym{wbf}{WBF}{Wired Bias Function}
\newacronym{wf}{WF}{Waterfilling}
\newacronym{wg}{WG}{Working Group}
\newacronym{wi}{WI}{Wireless InSite}
\newacronym{wlan}{WLAN}{Wireless Local Area Network}
\newacronym{osm}{OSM}{Open Source \gls{nfv} Management and Orchestration}
\newacronym{pnf}{PNF}{Physical Network Function}
\newacronym{mtc}{MTC}{Machine-type Communications}
\newacronym{mns}{MnS}{Management Services}
\newacronym{ves}{VES}{\gls{vnf} Event Stream}
\newacronym{ei}{EI}{Enrichment Information}
\newacronym{fh}{FH}{Fronthaul}
\newacronym{fft}{FFT}{Fast Fourier Transform}
\newacronym{laa}{LAA}{Licensed-Assisted Access}
\newacronym{plfs}{PLFS}{Physical Layer Frequency Signals}
\newacronym{ptp}{PTP}{Precision Time Protocol}
\newacronym{cbrs}{CBRS}{Citizen Broadband Radio Service}
\newacronym{otic}{OTIC}{Open Testing and Integration Center}
\newacronym{sba}{SBA}{Service-Based Architecture}
\newacronym{cif}{CI}{cyberinfrastructure}
\newacronym{sonic}{SONiC}{Software for Open Networking in the Cloud}
\newacronym{ocp}{OCP}{Open Compute Project}
\newacronym{snmp}{SNMP}{Simple Network Management Protocol}
\newacronym{raid}{RAID}{redundant array of independent disks}
\newacronym{nfs}{NFS}{Network File Storage}
\newacronym{ci}{CI}{Continuous Integration}
\newacronym{cd}{CD}{Continuous Deployment}
\newacronym{dtn}{DTN}{Data Transfer Node}

\newacronym{dns}{DNS}{Domain Name Service}
\newacronym{nrpe}{NRPE}{Nagios Remote Plugin Executor}
\newacronym{ldap}{LDAP}{Lightweight Directory Access Protocol}
\newacronym{lan}{LAN}{Local Area Network}
\newacronym{vlan}{VLAN}{Virtual LAN}

\newacronym{ipmi}{IPMI}{Intelligent Platform Management Interface}
\newacronym{tor}{ToR}{Top-of-the-Rack}
\newacronym{lmn}{LMN}{Large Memory Node}
\newacronym{bgp}{BGP}{Border Gateway Protocol}
\newacronym{dhcp}{DHCP}{Dynamic Host Configuration Protocol}
\newacronym{vrf}{VRF}{Virtual Routing and Forwarding}
\newacronym{vpn}{VPN}{Virtual Private Network}
\newacronym{rma}{RMA}{Return Merchandise Authorization}
\newacronym{hpc}{HPC}{High Performance Compute}

\newacronym{nu}{NU}{Northeastern University}
\newacronym{asic}{ASIC}{Application-specific Integrated Circuit}
\newacronym{rdma}{RDMA}{Remote Direct Memory Access}
\newacronym{roce}{RoCE}{RDMA over Converged Ethernet}
\newacronym{ovs}{OVS}{Open vSwitch}
\newacronym{frr}{FRR}{Free Range Routing}
\newacronym{ups}{UPS}{Uninterruptible Power Supply}

\newacronym{ntia}{NTIA}{National Telecommunications and Information Administration}
\newacronym{pii}{PII}{Personal and Identifiable Information}
\newacronym{irb}{IRB}{Institutional Review Board}
\newacronym{doi}{DOI}{Digital Object Identifier}

\newacronym{sdo}{SDO}{Standard-Development Organization}
\newacronym{ose}{OSE}{Open Source Ecosystem}
\newacronym{osc}{OSC}{O-RAN Software Community}
\newacronym{dop}{DOP}{Director of Operations}
\newacronym{pm}{PM}{Program Manager}
\newacronym{excom}{EXCOM}{Executive Committee}
\newacronym{iiot}{IIoT}{Industrial \gls{iot}}
\newacronym{lf}{LF}{Linux Foundation}

\newacronym{wiot}{WIoT}{Institute for the Wireless Internet of Things}
\newacronym{rl}{RL}{Reinforcement Learning}
\newacronym{drl}{DRL}{Deep Reinforcement Learning}

\newacronym{nofo}{NOFO}{Notice of Funding Opportunity}

\newacronym{onr}{ONR}{Office of Naval Research}
\newacronym{afosr}{AFOSR}{Air Force Office of Scientific Research}
\newacronym{afrl}{AFRL}{Air Force Research Laboratory}
\newacronym{arl}{ARL}{Army Research Laboratory}

\newacronym{arc}{ARC}{Aerial Research Cloud}
\newacronym{cast}{CaST}{Channel emulation scenario generator and Sounder Toolchain}

\newacronym{mno}{MNO}{Mobile Network Operator}
\newacronym{ct}{CT}{Continuous Testing}
\newacronym{oci}{OCI}{Open Container Initiative}

\newacronym{xai}{XAI}{Explainable AI}
\newacronym{esc}{ESC}{Environmental Sensing Capability}
\newacronym{sas}{SAS}{Spectrum Access System}

\newacronym{rem}{REM}{Random Ensemble Mixture}
\newacronym{ns3}{ns-3}{Network Simulator 3}
\newacronym{ngap}{NGAP}{Next Generation Application Protocol}
\newacronym{vue}{VUE}{Victim UE}

\usepackage[table]{xcolor}
\usepackage{colortbl}         
\usepackage{pifont}
\usepackage{soul} 
\usepackage{booktabs,array}
\usepackage{tabularx}
   
\definecolor{lightorange}{RGB}{255,180,80}
\newcommand{\cmark}{\textcolor{green!60!black}{\Large\ding{51}}}
\newcommand{\xmark}{\textcolor{red!70!black}{\Large\ding{55}}}
\newcommand{\amark}{\textcolor{lightorange}{\Large\ding{51}}}

\newcommand{\MinPts}{\mathit{MinPts }}
\newcommand{\epsrssi}{\epsilon_{\mathrm{RSSI }}}
\newcommand{\epsta}{\epsilon_{\mathrm{TA }}}
\newcommand{\name}{StormShield\xspace}

\usepackage{makecell}

\usepackage{float}

\usepackage[ruled,vlined,noend,linesnumbered]{algorithm2e} 

\usepackage{subcaption}

\makeatletter
\def\bstctlcite{\@ifnextchar[{\@bstctlcite}{\@bstctlcite[@auxout]}}
\def\@bstctlcite[#1]#2{\@bsphack
  \@for\@citeb:=#2\do{%
    \edef\@citeb{\expandafter\@firstofone\@citeb}%
    \if@filesw\immediate\write\csname #1\endcsname{\string\citation{\@citeb}}\fi}%
  \@esphack}
\makeatother 

\ccsdesc[500]{Networks~Denial-of-service attacks}

\copyrightyear{2026}
\acmYear{2026}
\setcopyright{cc}
\setcctype{by}
\acmConference[WiSec '26]{Proceedings of the 19th ACM Conference on Security and Privacy in Wireless and Mobile Networks}{June 30-July 03, 2026}{Saarbrücken, Germany}
\acmBooktitle{Proceedings of the 19th ACM Conference on Security and Privacy in Wireless and Mobile Networks (WiSec '26), June 30-July 03, 2026, Saarbrücken, Germany}
\acmDOI{10.1145/3765613.3811685}
\acmISBN{979-8-4007-2201-1/2026/06}

\newif\ifexttikz
\exttikzfalse
\ifexttikz
\else
\usepackage{tikzpagenodes,etoolbox}
\usetikzlibrary{calc}
\usepackage[contents={}]{background}
\AddEverypageHook{%
\ifnumequal{\thepage}{1}{%
    \tikz[remember picture,overlay]{%
        \node[draw,
        minimum width=1.03\textwidth,
        text width=1.02\textwidth,
        font=\footnotesize
        ]
        at ($(current page header area) - (0,-5pt)$)
        {%
        This paper has been accepted for publication on 19th ACM Conference on Security and Privacy in Wireless and Mobile Networks (ACM WiSec 2026). This is the author’s accepted version of the article. The final version published by ACM is
        N. Giustini, A. Lacava, L. Bonati, S. Maxenti, M. Polese, T. Melodia and F. Cuomo, "StormShield: Fingerprint-Based Detection and Mitigation of RRC Signaling Storms in O-RAN 5G RANs," 19th ACM Conference on Security and Privacy in Wireless and Mobile Networks (ACM WiSec 2026), Saarbrücken, Germany, pp. 1-11, doi: 10.1145/3765613.3811685, 2026.
        };
        \node[draw,
        minimum width=1.03\textwidth,
        text width=1.02\textwidth,
        font=\footnotesize
        ]
        at ($(current page footer area) - (0,20pt)$)
        {%
        ©2026 ACM. Personal use of this material is permitted. Permission from ACM must be obtained for all other uses, in any current or future media, including reprinting/republishing this material for advertising or promotional purposes, creating new collective works, for resale or redistribution to servers or lists, or reuse of any copyrighted component of this work in other works.
        };
    }%
}{}
}
\fi

\begin{document}

\title{StormShield: Fingerprint-Based Detection and Mitigation of RRC Signaling Storms in O-RAN 5G RANs}

\author{Noemi Giustini}
\affiliation{%
  \institution{Northeastern University}
  \city{Boston}
  \state{Massachusetts}
  \country{USA}}
\email{giustini.n@northeastern.edu}

\author{Andrea Lacava}
\affiliation{%
  \institution{Northeastern University}
  \city{Boston}
  \state{Massachusetts}
  \country{USA}}
\email{a.lacava@northeastern.edu}

\author{Leonardo Bonati}
\affiliation{%
  \institution{Northeastern University}
  \city{Boston}
  \state{Massachusetts}
  \country{USA}}
\email{l.bonati@northeastern.edu}

\author{Stefano Maxenti}
\affiliation{%
  \institution{Northeastern University}
  \city{Boston}
  \state{Massachusetts}
  \country{USA}}
\email{maxenti.s@northeastern.edu}

\author{Michele Polese}
\affiliation{%
  \institution{Northeastern University}
  \city{Boston}
  \state{Massachusetts}
  \country{USA}}
\email{m.polese@northeastern.edu}

\author{Tommaso Melodia}
\affiliation{%
  \institution{Northeastern University}
  \city{Boston}
  \state{Massachusetts}
  \country{USA}}
\email{t.melodia@northeastern.edu}

\author{Francesca Cuomo}
\affiliation{%
  \institution{Sapienza University of Rome}
  \city{Rome}
  \country{Italy}}
\email{francesca.cuomo@uniroma1.it}

\glsunset{usrp}

\begin{abstract}
5G networks provide low-latency, high throughput, and massive connectivity, yet the control plane remains exposed to several security threats. 
Among the most common and impactful threats are \gls{dos} attacks, with \gls{rrc} signaling storms being particularly effective and difficult to mitigate. In this attack, a malicious \gls{ue} aims to exhaust \gls{gnb} resources, preventing legitimate \glspl{ue} from establishing a connection.
Existing defenses are typically limited to detection, only evaluated through numerical simulations, and  cannot discern between high-load network conditions and attacks. Most of them also assume static setups and do not take mobility into account.
In this paper, we first evaluate the feasibility of the signaling storm attack by using the \gls{oai} 5G protocol stack.
Then, we propose \name, a signaling storm attack detection and mitigation technique implemented as an xApp on an O-RAN \gls{nearrt} \gls{ric}. It fingerprints and blocks \glspl{mue} before \gls{gnb} resources are exhausted.
We prototyped our solution on an \gls{ota} testbed with \gls{oai}, NVIDIA Aerial, and two different \gls{gnb} setups. The first one leverages an USRP X410 \gls{sdr} with 8.1 functional split; the second a commercial Foxconn \gls{ru} with 7.2 functional split. 
Our experimental evaluation demonstrates that \name effectively prevents \gls{gnb} resource exhaustion, identifying and blocking \glspl{mue} with an average detection accuracy of \textcolor{black}{97.6\%} within 106.5\,ms from the beginning of the attack.

\end{abstract}

\keywords{O-RAN, 5G, Security, RRC Signaling Storm, Denial of Service.}

\renewcommand{\shortauthors}{Noemi Giustini et al.}

\maketitle
\glsresetall
\glsunset{usrp}
\glsunset{cpu}
\glsunset{gpu}
\glsunset{ram}
\glsunset{nr}

\section{Introduction}
\begin{figure}[h]
  \centering
  \begin{subfigure}[b]{0.23\textwidth}
    \centering
    \includegraphics[width=.9\linewidth]{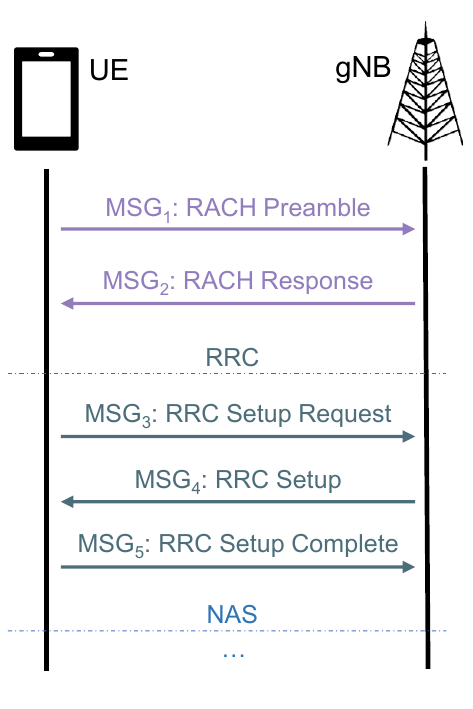}
    \caption{Normal UE Behavior}
    \label{fig:UE_Normal_Behaviour}
  \end{subfigure}
  \hfill
  \begin{subfigure}[b]{0.23\textwidth}
    \centering
    \includegraphics[width=.9\linewidth]{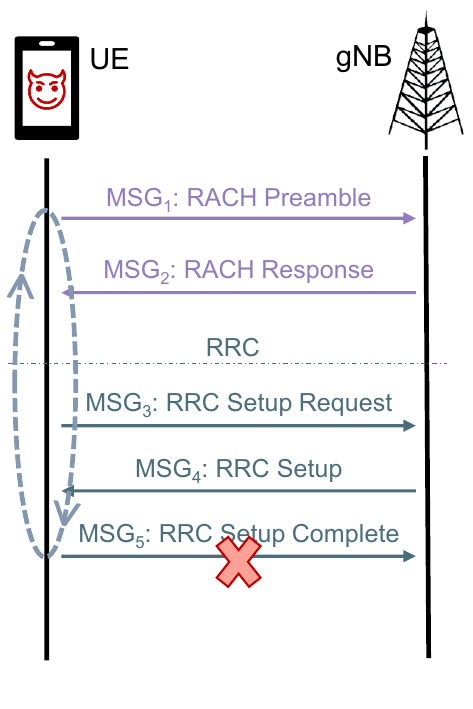}
    \caption{Malicious UE Behavior}
    \label{fig:UE_Malicious_Behaviour}
  \end{subfigure}
  \caption{Comparison between Normal UE and MUE Behavior}
   \Description{Comparison between normal and MUE behavior}
  \label{fig:UE_Behaviours}
\end{figure}

As the \gls{5g} of cellular networks scales up to support enhanced mobile broadband, ultra-reliable low-latency connections, and massive machine-type communications, ensuring high availability in the protocol stack control-plane emerges as a major security concern. 
Both user and control planes are exposed to \gls{dos} attacks that can degrade or interrupt network service~\cite{kim2019touching, liao2022development, feliana2024evaluation}, but control-plane attacks are more concerning, as they can be carried out by unauthenticated users.
Among these attacks, the \gls{rrc} signaling storm constitutes a relevant risk for the control-plane. 
In this attack, an adversarial \gls{ue}, which we will refer to as \gls{mue} throughout the rest of the paper, repeatedly generates \gls{rrc} Setup Requests with different \glspl{rnti} without sending the \gls{rrc} Setup Complete message~\cite{oran-wg1-use-cases}. In this way, the \gls{mue} avoids authenticating to the network but forces the \gls{gnb} to allocate resources for its session (as shown in Fig.~\ref{fig:UE_Malicious_Behaviour}).
This malicious behavior depletes control-plane capacity and prevents legitimate connections~\cite{nguyen2025rrc}.

The detection of these attacks is challenging because the initial observable behavior mimics genuine high-load events in which the \gls{gnb} observes bursts of \gls{rrc} Setup Requests within short time windows.
This makes rate-based thresholds~\cite{wen20245g} prone to false alarms.
Mitigation is even more challenging and far less explored in the literature. Since \glspl{mue} never complete the authentication procedure, their temporary identifiers (e.g., \gls{rnti}) change at every new connection attempt. This makes it difficult for the \gls{gnb} to attribute subsequent connection attempts to specific \glspl{ue}, and, thus, to implement per-\gls{ue} blocking policies.
Consequently, effective mitigation must rely on device- or transmission-level fingerprints (e.g., \gls{ta})~\cite{10226043}.

We leverage O-RAN as an architectural enabler for addressing the above challenges and design our solution. By exposing standardized, open interfaces toward a \gls{nearrt} \gls{ric}, O-RAN provides the observability and control loops needed to collect fine-grained \glspl{kpm} and react rapidly in case of an attack~\cite{bonati2020open}.
Specifically, we introduce \name, a \gls{nearrt} detection and mitigation algorithm for \gls{rrc} signaling storm attacks that clusters \glspl{ue} by using multiple radio-level indicators---\gls{rssi} and \gls{ta}---and access pattern features (e.g., per-source burst of connection attempts and request-to-completion ratios) to build a robust \glspl{mue} fingerprint.
The combination of multiple indicators is crucial, as existing defenses---mainly detection-only or volume-based---struggle to separate genuine high load from attacks, and \gls{ta} alone is insufficient in dense networks where \glspl{ue} share similar values for this parameter.
Differently from prior work, \name (i) implements an effective attack detection and mitigation, rejecting connection attempts from \glspl{ue} flagged as malicious before \gls{gnb} resources saturate; (ii) discerns high-load vs. attack scenarios; and (iii) remains robust in challenging environments, including multiple simultaneous attackers and mobility.

We prototype \name on an \gls{ota} testbed with programmable 5G protocol stacks (OAI, NVIDIA Aerial)~\cite{kaltenberger2025driving,kelkar2021nvidia} and two different \gls{gnb} setups. In the first, \gls{oai} drives an USRP X410 \gls{sdr} with 8.1 functional split to perform signal transmission and reception. In the second setup, instead, the \gls{oai} protocol stack connects to the GPU-accelerated NVIDIA ARC solution, which implements the lower-layer functionalities of the \gls{gnb} protocol stack, and connects to a commercial \gls{ru} from Foxconn with 7.2 split. We prototype the malicious \gls{ue} by customizing the \gls{rrc} state machine in the OAI software \gls{ue}, and use commercial smartphones as legitimate users.
Experimental results in the different setups demonstrate that \name is effective in detecting and mitigating signaling storm attacks before the \gls{gnb} resources are depleted, achieving a \textcolor{black}{97.6\%} average accuracy. 
Our key contributions are the following:
\begin{itemize}
    \item \textbf{Attack feasibility in realistic 5G setups -} We reproduce the \gls{rrc} signaling storm attack over different \gls{ota} testbed setups, including effects such as mobility and multiple concurrent attackers.

    \item \textbf{Fingerprint-based detection of signaling storms -} We design a detection algorithm that combines \gls{rrc}-layer statistics with spatial fingerprinting. This allows the system to distinguish signaling storm attacks from benign high-load network conditions even in challenging indoor scenarios where distance-based features alone are not sufficient.

    \item \textbf{Pre-authentication mitigation via fingerprint blocklisting -} We introduce a mitigation algorithm that operates before authentication and without relying on stable user identities. It maintains an adaptive block list of malicious fingerprints with aging timeouts that reinforce blocking for persistent attackers, while automatically expiring stale entries to avoid long-term exclusion of benign \glspl{ue}. We demonstrate the effectiveness of this technique in the presence of multiple attackers, and both static and mobile devices.

    \item \textbf{O-RAN-based closed control loop implementation -} We deploy \name in an xApp over the \gls{nearrt} \gls{ric}, enabling closed-loop detection and mitigation without modifying the \gls{ran} data plane. We extend the O-RAN E2 data-exposure capabilities with a new security-oriented \gls{sm}, implemented according to O-RAN use case~\cite{oran-wg1-use-cases}, that defines telemetry and control semantics to obtain the \glspl{kpm} required by \name.
\end{itemize}

The remainder of this paper is organized as follows. Section~\ref{sec:soa} reviews different types of \gls{dos} attacks in O-RAN and the detection and mitigation solutions proposed in the literature. 
\textcolor{black}{Section~\ref{sec:oran-arch} provides a brief technical background on the Open RAN architecture and the integration of xApps within the \gls{nearrt} \gls{ric}, highlighting the main interfaces leveraged in this work.}
Section~\ref{sec:attack} introduces our threat model and characterizes the signaling storm attack. Section~\ref{sec:logic} focuses on the detection and mitigation logic and details the implementation of the \gls{sm} for \glspl{kpm} collection. 
Section~\ref{sec:architecture} describes the \name prototype and the experimental setup used in our evaluation. 
Section~\ref{sec:results} discusses our experimental results under different setups and conditions. Finally, Section~\ref{sec:conclusions} concludes our work and discusses possible future directions.

\section{Related Work}
\label{sec:soa}

In this section, we first survey \gls{dos} attacks in cellular networks (Sec.~\ref{sec:dos-cellular}).
Then, we focus on \gls{rrc} signaling storm detection and mitigation methods used in the literature (Sec.~\ref{sec:sign-storm-det-mit}).
Finally, we outline key elements of the Open \gls{ran} architecture relevant to security and data exposure (Sec.~\ref{sec:oran-arch}).

\subsection{Denial of Service Attacks in Cellular Networks}
\label{sec:dos-cellular}
\begin{table*}[t]
\centering
\small
\setlength{\tabcolsep}{6pt}
\renewcommand{\arraystretch}{1.2}
\setlength\abovecaptionskip{2pt}
\caption{Literature Excerpt on Signaling Storm Attacks and Comparison of Existing Solutions against \name}
\begin{tabular}{
>{\raggedright\arraybackslash}p{3cm}
>{\raggedright\arraybackslash}p{1.8cm}
>{\raggedright\arraybackslash}p{1cm}
>{\raggedright\arraybackslash}p{1.3cm}
>{\raggedright\arraybackslash}p{1.3cm}
>{\raggedright\arraybackslash}p{1.8cm}
>{\raggedright\arraybackslash}p{4.5cm}
}
\toprule
\textbf{Paper} &
\textbf{Setup} &
\textbf{Stack} &
\textbf{Detection} &
\textbf{Mitigation} &
\textbf{UE Mobility} &
\textbf{Limitations} \\
\midrule
Kim et al.~\cite{kim2019touching} & OTA & LTE & \xmark & \xmark & Static & No detection/mitigation presented \\
Wen et al.~\cite{wen20245g} & OTA + Simulator & 5G & \cmark & \xmark & Static & Detection cannot distinguish between attack and high demand \\
Nguyen et al.~\cite{nguyen2025rrc} & Simulator & 5G & \cmark & \xmark & Static & Detection relies only on static thresholds \\
Hoffmann et al.~\cite{10226043} & Simulator & 5G & \cmark & \amark (partially) & Static & TA cannot distinguish co-located UEs (e.g., indoor) \\
\name (This Work) & Different RUs as OTA gNBs & 5G & \cmark & \cmark & Static/Mobile (MUE/VUE)& -- \\
\bottomrule
\end{tabular}
\label{tab:soa}
\end{table*}

Numerous types of \gls{dos} attacks have been explored in recent literature on \gls{5g} cellular networks, highlighting how adversaries can exploit protocol vulnerabilities and resource constraints to degrade or even disrupt service availability. One example is presented in~\cite{benzaid2024federated}, where the authors introduce TenaxDoS, a detection framework based on deep learning and federated continual learning, and deployed in an O-RAN \gls{5g} mobile network. This system focuses on identifying distributed \gls{dos} attacks targeting the traffic volume (e.g., bandwidth saturation), the protocol stack (e.g., handshake abuse), and application-layer aspects (e.g., resource draining).
The authors of~\cite{liao2022development} design a tool capable of launching control-plane \gls{dos} attacks via the O-RAN fronthaul interface. The tool generates \gls{rt} control messages over Ethernet using the eCPRI protocol and can spoof MAC addresses to appear as O-\glspl{ru} or broadcast entities.
Another work~\cite{feliana2024evaluation} demonstrates both control and user plane \gls{dos} attacks using a custom-built evaluation tool. The authors assess the impact on throughput, \gls{bler}, and the state evolution of O-\gls{du}/\gls{ru}.

Further \gls{dos} vectors are described in~\cite{klementendless,hung2024security}, both targeting the \gls{ric} E2 subscription mechanism via a malicious xApp that exploits weak access control and input validation to disrupt the \gls{nearrt} \gls{ric} by sending numerous crafted or malformed subscription requests.
The former work floods the \gls{nearrt} \gls{ric} with legitimate-looking E2 requests to degrade service. The latter demonstrates that malformed messages can crash the E2 termination component of the \gls{ric}.

The work in~\cite{janzen2024oh} reveals vulnerabilities that allow adversaries to cause \gls{dos} on the \gls{ru}. 
Attackers with limited access can interrupt data transmission by changing \gls{rf} parameters, misconfigure VLANs to block access, or repeatedly reboot the \gls{ru} to make it unavailable.

In~\cite{kim2019touching}, the authors describe \gls{dos} attacks such as the Blind \gls{dos} and the \gls{bts} resource depletion attacks. In the former, an adversary spoofs the \gls{s-tmsi} of the victim \gls{ue} to initiate \gls{rrc} connections on its behalf.
In the latter, an adversary floods the \gls{gnb} with \gls{rrc} connection requests without completing authentication, exhausting the maximum number of allowed \gls{rrc} connections and blocking legitimate \glspl{ue} from connecting to the \gls{gnb}.
Our work focuses specifically on this type of attack, often referred to as a \gls{rrc} Signaling Storm.
Differently from the above literature works---and beyond reproducing and evaluating the attack on a real \gls{5g} \gls{ota} testbed with mobility and multiple \glspl{mue}---we propose a fingerprint-based detection and mitigation logic that blocks malicious connection attempts, prevents them from exhausting the \gls{gnb} resources, while providing service for legitimate \glspl{ue}.

\subsection{Signaling Storm Detection and Mitigation}
\label{sec:sign-storm-det-mit}
These attacks are difficult to detect due to changing attacker identifiers and the similarity of the traffic pattern to that of a high-load network. Despite their impact, the literature lacks effective mitigation strategies, because attackers never complete the authentication phase and continuously change their temporary identifiers. 
Table~\ref{tab:soa} summarizes the main literature approaches on this topic,
and highlights how our work differs in terms of setup, attack detection and mitigation capabilities.

In~\cite{tsourdinis2024ai}, an Anomaly Traffic Detector xApp applies \gls{rt} traffic classification using machine learning. It identifies anomalous users analyzing packets and using a random forest. If the \gls{ue} anomaly ratio reaches 100\%, the xApp triggers an \gls{rrc} \gls{ue} connection release, causing the \gls{ue} to disconnect from the network. However, this method assumes the attacker has completed the authentication process, which is not the case in signaling storm scenarios where the attacker avoids authentication altogether.

Another work~\cite{wen20245g} introduces a framework that detects several types of cellular protocol exploits in \gls{lte} networks, including attacks that exploit the \gls{enb} (e.g., \gls{bts} resource depletion), the \gls{mme} of the core network, and even the \glspl{ue} themselves. Considering the resource depletion attack, its detection method relies on the fact that fake \gls{rrc} connections---dubbed ``transient \glspl{ue}''---are typically short-lived due to a missing \gls{nas} authentication response. 
If the number of transient \glspl{ue} exceeds a threshold, the attack is detected. 
However, this approach cannot reliably distinguish an attack from a legitimate high-load condition of the network.

In~\cite{wen20246g}, the authors propose 6G-XSec, a framework that combines unsupervised deep learning with \glspl{llm} to monitor and interpret network anomalies, including \gls{bts} resource depletion attacks. The system first extends O-RAN with ``MOBIFLOW'' to collect telemetry data including encoded RRC/NAS messages and parameters such as \gls{ue} identifiers and state. The collected time-series traces are then fed to an unsupervised anomaly-detection xApp (namely, ``MOBIWATCH''), trained on benign traffic only. In the case of \gls{bts} \gls{dos} attacks, the detector learns to flag sequences with abnormal patterns of \gls{rrc}/\gls{nas} messages and device identifiers, such as rapid bursts of uncompleted connection attempts with changing \gls{rnti} values. An \gls{llm}-based xApp is subsequently used to classify and explain the detected anomalies. A limitation of this approach is that benign but unusual conditions (e.g., multiple retransmissions or connection failures) can produce similar anomalous patterns, making it difficult to reliably separate them from malicious activity.

In~\cite{nguyen2025rrc}, a threshold-based detection technique is proposed. This uses \gls{rrc}-layer metrics like the number of MSG3 messages (\gls{rrc} Connection Request), and the MSG5-to-MSG3 and MSG5-to-MSG4 ratios, where MSG4 and MSG5 correspond to the \gls{rrc} Connection Setup and \gls{rrc} Connection Setup Complete messages, respectively. A surge in MSG3 indicates abnormal behavior, while a drop in the ratios can help distinguish an attack from high-load conditions, since in high-load conditions most \glspl{ue} that receive MSG4 still respond with MSG5, whereas this does not happen during an attack. However, this approach is limited by its reliance only on fixed thresholds, which are not adaptive to different network conditions and may lead to false positives.

Overall, the above methods mostly focus on detection rather than mitigation, as we instead do in this work.
A partially implemented mitigation solution is proposed in~\cite{10226043}, where a dedicated xApp uses \gls{ta} to fingerprint devices early in the connection process. Since \gls{ta} correlates with the physical distance between the \gls{ue} and base station, this parameter helps to identify and block suspicious devices before they authenticate. However, this method becomes unreliable in indoor or dense scenarios, where \glspl{ue} are close, as \gls{ta} lacks the granularity to differentiate between them.
Additionally, most of the above solutions have been tested in simulations or in limited testbeds, often involving only a single stationary attacker. By contrast, our logic handles single or multiple, stationary or moving attackers, and was validated on an \gls{ota} testbed with different setups and configurations.

\section{\textcolor{black}{A Primer on the Open RAN Architecture}}
\label{sec:oran-arch}

The Open \gls{ran} architecture, and its embodiment in O-RAN, acts as a key enabler of closed-loop control for addressing \gls{5g} \gls{ran} performance and security issues.
O-RAN builds on the \gls{3gpp} \gls{gnb} disaggregation into \gls{cu}, \gls{du}, and \gls{ru} elements. These elements can be virtualized on the O-Cloud infrastructure and interconnected via open and standardized interfaces~\cite{polese2023understanding}. O-RAN introduces two \glspl{ric} acting at different time scales. The \gls{nonrt} \gls{ric}, integrated within the \gls{smo}, operates closed-loop control of the \gls{ran} at time scales above $1$\,s; the \gls{nearrt} \gls{ric}, deployed at the edge, operates at time scales between $10$\,ms and $1$\,s. \textcolor{black}{Both \glspl{ric} host intelligent \gls{ai}/\gls{ml} applications.} The \gls{nearrt} \gls{ric} hosts xApps that subscribe to telemetry (e.g., \glspl{kpm}) exposed over the O-RAN E2 interface and enforce control actions on E2 nodes (i.e., \gls{cu}, \gls{du}, \textcolor{black}{and legacy \gls{enb} supporting this interface}). xApps realize data-driven closed-control loops that can promptly react to network anomalies, such as the \gls{dos} attacks \textcolor{black}{we focus on in this work}.
\textcolor{black}{The \gls{nonrt} \gls{ric}, instead, hosts rApps that receive telemetry at a larger granularity over the O-RAN O1 interface from the remaining Open \gls{ran} elements (i.e., \gls{nearrt} \gls{ric}, \gls{cu}, \gls{du}, \gls{ru}, and \textcolor{black}{legacy \gls{enb} supporting this interface}). rApps push high-level network policies to the \gls{nearrt} \gls{ric} via the O-RAN A1 interface that connects the two \glspl{ric}.
Finally, the \gls{nonrt} \gls{ric} connects to the O-Cloud infrastructure via the O-RAN O2 interface to carry monitoring operations, as well as programmatic provisioning, management, and update of network functions.}

In this work, we leverage the openness and control capabilities of the Open \gls{ran} architecture to implement \name’s attack detection and mitigation logic, as discussed in Sec.~\ref{sec:logic}.

\section{Attack Modeling}
\label{sec:attack}

The \gls{5g} connection establishment procedure, as defined in the \gls{3gpp} specifications~\cite{etsi_ts_138331_v1860}, enables a \gls{ue} to access the network for services such as data and voice communication.
As illustrated in Fig.~\ref{fig:UE_Normal_Behaviour}, the procedure begins with cell search and synchronization. The \gls{ue} then acquires the necessary uplink access parameters and initiates the \gls{rach} procedure by transmitting a random preamble (MSG1) to the \gls{gnb}. At this point, the \gls{ue} also starts a timer (namely, timer T300), which sets the time limit for completing the connection setup. If the timer expires, the \gls{ue} considers the attempt failed and restarts the process.

Upon receiving the preamble, the \gls{gnb} detects the transmission and responds with a Random Access Response (MSG2), which includes a \gls{ta}, an uplink grant, and a temporary C-RNTI. This allows the \gls{ue} to proceed with MSG3, an \gls{rrc} Setup Request message that carries the \gls{ue} identity and the establishment cause.
The \gls{gnb}, then, replies with an \gls{rrc} Setup (MSG4), which defines radio bearer configurations and resolves contention. Upon receiving this message, the \gls{ue} stops the T300 timer and responds with an \gls{rrc} Setup Complete message (MSG5), which includes a \gls{nas} Registration Request. Finally, the \gls{gnb} selects an appropriate \gls{amf} and assigns a \gls{ngap} identifier to the \gls{ue}. This completes the \gls{ue} connection process.

In contrast, in the case of a \gls{rrc} signaling storm attack, the \gls{mue} exploits this mechanism as illustrated in Fig.~\ref{fig:UE_Malicious_Behaviour}. 
The attacker initiates the connection process by sending MSG1 through MSG3 but intentionally avoids completing the setup by not replying with MSG5 once it receives MSG4 from the \gls{gnb}.
Despite not completing the connection, the \gls{gnb} still allocates radio resources and context to the \gls{ue} based on MSG3, and starts a timer while waiting for MSG5~\cite{nguyen2025rrc}.
Once MSG3 is decoded, the \gls{gnb} creates an \gls{rrc} \gls{ue} context, assigning an internal \gls{rrc} \gls{ue} identifier, binding the temporary \gls{rnti}, and allocating the associated control-plane state (e.g., security capabilities, bearer configuration and timers). 
Even though the \gls{mue} never completes the \gls{rrc} procedure, this per-\gls{ue} context remains stored in the \gls{ue} context maintained by the \gls{gnb}, needlessly occupying \gls{rrc} context resources.
Under a signaling storm attack, many of such unauthenticated contexts accumulate, rapidly exhausting the pool of \gls{rrc} contexts available at the \gls{gnb} and preventing new attachments from legitimate \glspl{ue}.

\section{Attack Detection and Mitigation}
\label{sec:logic}

As discussed in Sec.~\ref{sec:attack}, \gls{rrc} signaling storm attacks are difficult to detect as they resemble high-load network conditions.
Moreover, their mitigation is also challenging, as \glspl{mue} are never assigned stable identifiers as they do not complete the authentication procedure.
In this section, we will detail how \name can effectively detect (Sec.~\ref{sec:detection_algorithm}) and mitigate (Sec.~\ref{sec:mitigation_algorithm}) this kind of attacks.

\begin{algorithm}[h]
\caption{Signaling Storm detection algorithm}
\label{alg:storm_detection}
\small
\While{true}{
  Receive $N_3, N_4, N_5$ and fingerprint history 
    $\mathcal{X} = \{(\mathrm{TA}_i,\mathrm{RSSI}_i)\}_{i=1}^{M}$\;
  $R_1 \gets N_5 / N_3$,\quad $R_2 \gets N_5 / N_4$\;

  \If{$N_3 > T_1$ \textbf{and} $R_1 < T_2$ \textbf{and} $R_2 < T_2$}{
    Run DBSCAN on 
    $\mathcal{X}$\;
    
    $\mathcal{F} \gets \emptyset$\;
    \ForEach{cluster $C_k$}{
      $P_k \gets |C_k| / M$\;
      \If{$P_k > T_3$}{
        Compute centroid $\boldsymbol{\mu}_k$ of $C_k$\;
        Add $\boldsymbol{\mu}_k$ to $\mathcal{F}$\;
      }
    }
    \textbf{ATTACK DETECTED:} output $\mathcal{F}$\;
  }
  \ElseIf{$N_3 > T_1$ \textbf{and} $R_1 \geq T_2$ \textbf{and} $R_2 \geq T_2$}{
    \textbf{HIGH-LOAD DETECTED}
  }
  \Else{
    \textbf{NORMAL LOAD}\;
  }
}
\end{algorithm}

\subsection{Detection Algorithm}
\label{sec:detection_algorithm}

The detection logic of \name is shown in Algorithm~\ref{alg:storm_detection}. At a high level, it operates on periodic measurements collected over a sliding time window of duration $W$. For each window, we consider the number of \gls{rrc} MSG3, MSG4, and MSG5 messages observed, denoted by $N_3$, $N_4$, and $N_5$, respectively, as well as \gls{ta} and \gls{rssi} metrics that we use to \textit{fingerprint} \glspl{ue}.
For each window, the detector monitors the amount of MSG3 and two ratios, similarly to what done in~\cite{nguyen2025rrc}:
\begin{align}
R_1 &= \frac{N_5}{N_3} & \qquad
R_2 &= \frac{N_5}{N_4}
\end{align}
A large number of MSG3 ($N_3 > T_1$) indicates that many connection attempts are being made. If this is accompanied by low completion ratios ($R_1 < T_2$ and $R_2 < T_2$), it suggests that a significant fraction of connection procedures do not reach MSG5, which is an indicator of a signaling storm attack rather than standard high-load conditions~\cite{nguyen2025rrc}, with $T_1$, $T_2$, and $T_3$ tunable thresholds in \name.
Indeed, under a signaling storm, many MSG3 attempts from the same \gls{mue} are expected to cluster under similar fingerprints, whereas, under benign high-load conditions, MSG3 messages are spread across many distinct \glspl{ue} and, thus, fingerprints.
\textcolor{black}{This design choice is important to reduce false positives in realistic scenarios of legitimate mass access (e.g., many \glspl{ue} attempting attachment concurrently when entering a new cell). In such cases, despite a temporary burst in MSG3, most procedures still complete successfully and progress beyond MSG5, implying high completion ratios. Conversely, during signaling storm attacks the system observes persistently low completion ratios for a dominant fingerprint cluster, since \glspl{mue} intentionally avoid completing the connection procedure.}

When this coarse-grained event is triggered, the \name algorithm refines the attack/no-attack decision by analyzing the spatial structure of the \gls{ue} fingerprint, and we interpret each pair as a point in a bi-dimensional space:
\begin{equation}
\label{eq:x}
    \mathbf{x}_i = (\mathrm{TA}_i,\mathrm{RSSI}_i) \in \mathbb{R}^2
\end{equation}
that corresponds to a single MSG3 $i$ received in the current time window, and stores the associated timing advance $\mathrm{TA}_i$ (which correlates with the distance between \gls{gnb} and \gls{ue}) and received signal strength indicator $\mathrm{RSSI}_i$.
We leverage Eq.~\eqref{eq:x} to build a \textit{fingerprint history}, implemented as a circular buffer of size~$M$ and defined as:
\begin{equation}
\label{eq:fingerprint-history}
    \mathcal{X} = \{ \mathbf{x}_i \}_{i=1}^{M}.
\end{equation}

\begin{figure}[ht]
  \centering
  \begin{subfigure}[b]{0.20\textwidth}
    \centering
    \includegraphics[width=\linewidth]{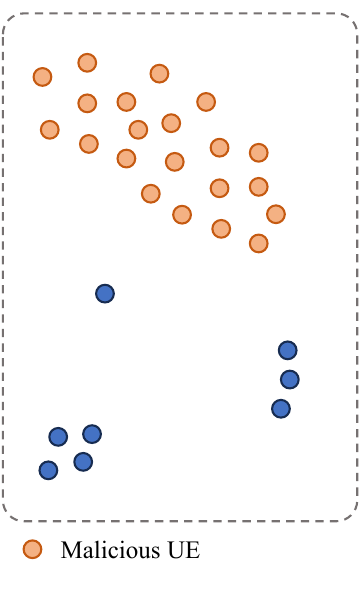}
    \caption{Attack Scenario}
    \label{fig:clustering_attack}
  \end{subfigure}
  \hfill
  \begin{subfigure}[b]{0.20\textwidth}
    \centering
    \includegraphics[width=\linewidth]{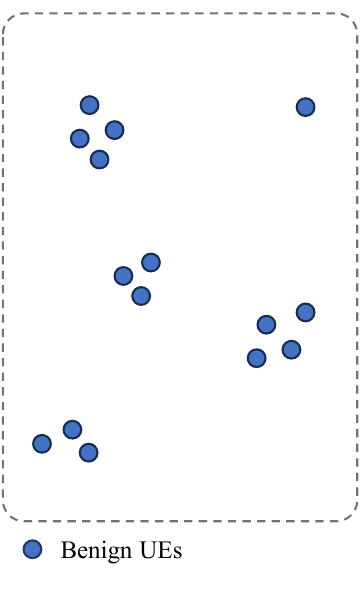}
    \caption{High-load Scenario}
    \label{fig:clustering_high_load}
  \end{subfigure}
  \vspace{-0.3cm}
  \caption{Clustering Results: Denser Clusters during an Attack}
  \Description{Clustering of attack vs. high-load scenarios}
\end{figure}

We apply the \gls{dbscan} clustering algorithm~\cite{ester1996density} to the fingerprint history of Eq.~\eqref{eq:fingerprint-history}.
Specifically, we use a normalized Euclidean distance $d_{i,j}(\mathbf{x}_i,\mathbf{x}_j)$ that combines deviations in timing advance and received power 
together with \gls{dbscan} parameters $(\varepsilon, \text{MinPts})$ such that points are considered neighbors if $d_{i,j}(\mathbf{x}_i,\mathbf{x}_j) \leq \varepsilon$, and a cluster must contain at least \text{MinPts} points. 
The rationale behind this is that a \gls{mue} that repeatedly sends MSG3 messages from the same location generates a dense cluster of very similar fingerprints in the $(\mathrm{TA},\mathrm{RSSI})$ space, whereas benign network high loads produce a more scattered pattern. 
\gls{dbscan} returns a set of clusters ${C_k}$, each with size $|C_k|$, where $k$ is the cluster index. For each cluster, we compute its relative density $P_k$ as
\begin{equation}
  P_k = \frac{|C_k|}{M}.
\end{equation}
A cluster is flagged as \emph{malicious} if $P_k$ exceeds a density threshold $T_3$.
This condition ensures that the cluster accounts for a significant portion of all MSG3s in the window, which is a strong indication that a single device is sending the majority of connection attempts, as shown in Fig.~\ref{fig:clustering_attack}. In contrast, during benign high-load scenarios (Fig.~\ref{fig:clustering_high_load}) MSG3s are distributed across many \glspl{ue} and no single cluster reaches a large occupancy of the fingerprint history buffer. 

The output of the detection algorithm is a set of \emph{malicious fingerprints}, represented by the cluster centroids

\begin{equation}
  \mathcal{F} = \{ \mu_k \}_{k \in \mathcal{K}},
\end{equation}
where $\boldsymbol{\mu}_k$ is the centroid of a malicious cluster $C_k$ and $\mathcal{K}$ is the index set of clusters that satisfy $P_k > T_3$. These fingerprints summarize the spatial signature of potential attackers and are passed to the mitigation algorithm, which we describe next.

\subsection{Mitigation Algorithm}
\label{sec:mitigation_algorithm}

As described in Algorithm~\ref{alg:storm_mitigation}, the \gls{gnb} maintains a \emph{block list} $\mathcal{B}$, where each entry $e \in \mathcal{B}$ is a tuple

\begin{equation}
  e = (\mu, c, \tau, t_{\mathrm{last}}),
\end{equation}

\noindent
where $\mu$ is the malicious fingerprint centroid received from the detector; $c$ is a match counter indicating how many times this fingerprint has matched incoming connection attempts; $\tau$ is a timeout that defines how long the fingerprint entry remains active without new matches; and $t_{\mathrm{last}}$ is the timestamp of the most recent match.

Whenever the detector produces a new set of malicious fingerprints $\mathcal{F}$, the mitigation logic updates the block list $\mathcal{B}$.
If a centroid $\boldsymbol{\mu} \in \mathcal{F}$ is already present in $\mathcal{B}$, its counter $c$ is incremented, $t_{\mathrm{last}}$ is updated, and the timeout $\tau$ is increased linearly (up to a maximum) to reinforce the block against persistent attackers. 
If $\boldsymbol{\mu}$ is new, a fresh entry is created with an initial counter and timeout.

For each new \gls{rrc} connection attempt (e.g., Setup or Reestablishment), the \gls{gnb} measures the current fingerprint $\mathbf{f} = (\mathrm{TA}, \mathrm{RSSI})$ and compares it with the active entries in $\mathcal{B}$. Before matching, the algorithm removes any entry whose inactivity exceeds its timeout, i.e., all $e$ such that $t_{\mathrm{now}} - t_{\mathrm{last}} \geq \tau$.
This aging mechanism is crucial, as, if an entry never expired, a benign \gls{ue} establishing a subsequent connection with similar transmission conditions could match an old malicious fingerprint and incorrectly be blocked.

\begin{figure*}[ht]
    \centering
    \begin{subfigure}[t]{0.475\textwidth}
        \centering
        \includegraphics[width=\textwidth]{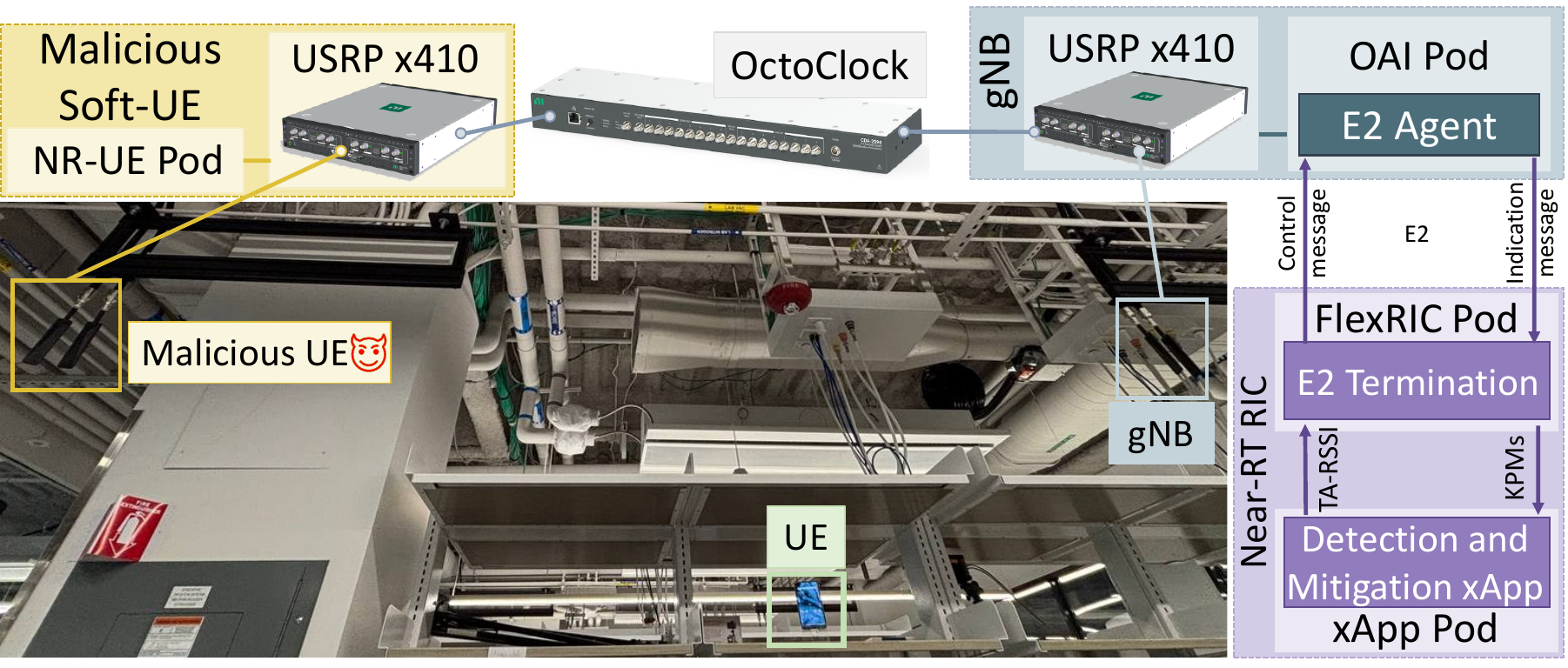}
        \caption{USRP X410 SDR}
        \label{fig:setup_usrp}
    \end{subfigure}
    \hfill
    \begin{subfigure}[t]{0.512\textwidth}
        \centering
        \includegraphics[width=\textwidth]{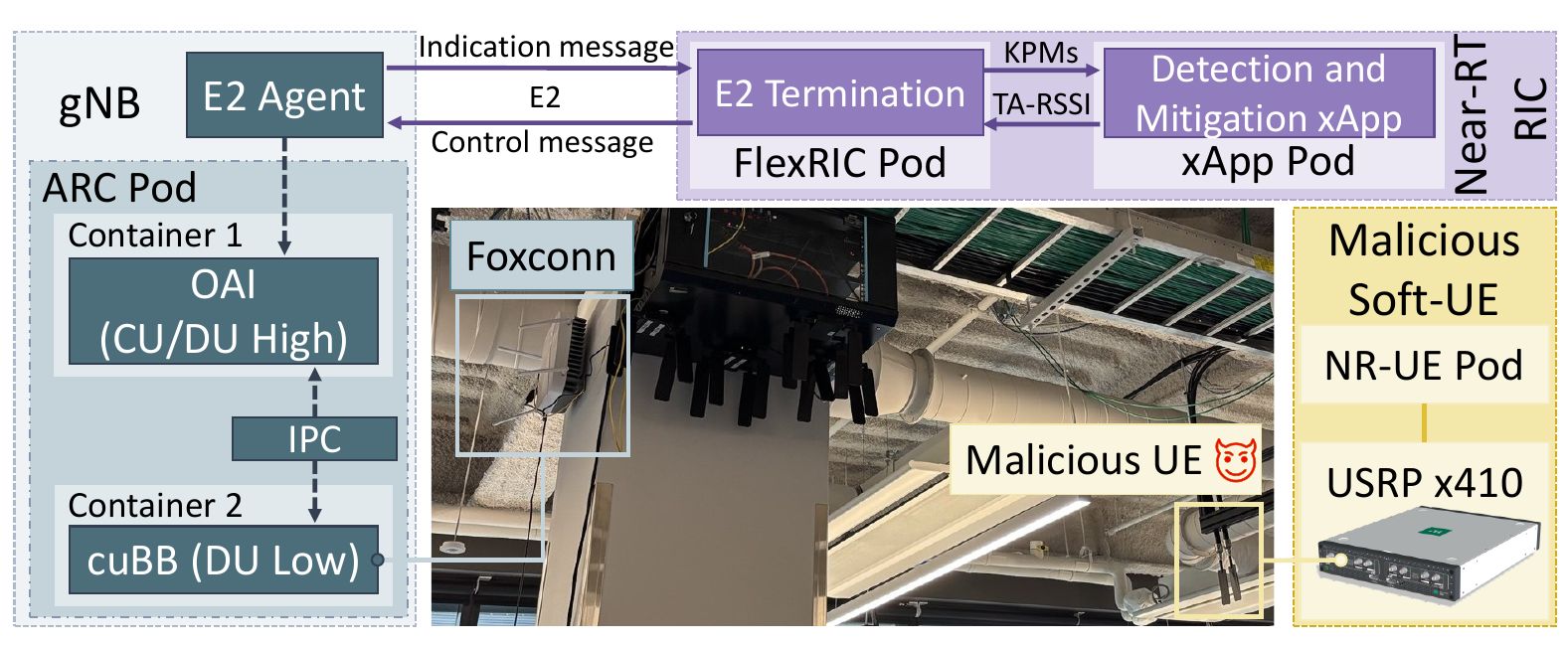}
        \caption{Foxconn RU}
        \label{fig:setup_foxconn}
    \end{subfigure}
    \vspace{-0.3cm}
    \caption{Architectures of the OTA Testbed Setups used to Prototype and Evaluate \name}
    \Description{Architectures of the OTA testbed setups used to prototype and evaluate \name}
    \label{fig:hw_setups}
\end{figure*}

For the remaining entries, the algorithm checks whether the current fingerprint $\mathbf{f} = (\mathrm{TA}, \mathrm{RSSI})$ is sufficiently close to any centroid $\boldsymbol{\mu} = (\mu_{\mathrm{TA}}, \mu_{\mathrm{RSSI}})$ in the block list.
$\epsta$ and $\epsrssi$ define the maximum allowed difference in \gls{ta} and \gls{rssi}, respectively, for two points to be considered as belonging to the same \gls{ue}.
A match occurs if the following two conditions are satisfied for some centroid $\boldsymbol{\mu}$:

\begin{align}
  |\mathrm{TA} - \mu_{\mathrm{TA}}| &\leq \epsta \\
  |\mathrm{RSSI} - \mu_{\mathrm{RSSI}}| &\leq \epsrssi,
\end{align}

\noindent
in which case, the connection attempt is rejected and the corresponding \gls{rrc} context dropped, thus preventing the attack from consuming additional resources. Otherwise, the connection proceeds as normal.

\begin{algorithm}[h]
\caption{Fingerprint-based Mitigation Algorithm}
\label{alg:storm_mitigation}
\small
Initialize block list $\mathcal{B} \gets \emptyset$\;
\While{true}{
  \If{new malicious fingerprint set $\mathcal{F}$ is received}{
    \ForEach{$\boldsymbol{\mu} \in \mathcal{F}$}{
      \If{$\boldsymbol{\mu}$ in $\mathcal{B}$}{
        Increment match counter $c$; update $t_{\mathrm{last}}$\;
        \If{$c \bmod k = 0$}{
          Increase timeout $\tau$ (capped at $\tau_{\max}$)\;
        }
      }
      \Else{
        Add entry $(\boldsymbol{\mu}, c=1, \tau=\tau_0, t_{\mathrm{last}}=t_{\mathrm{now}})$ to $\mathcal{B}$\;
      }
    }
  }
  \ForEach{incoming RRC connection attempt}{
    Measure current fingerprint $\mathbf{f} = (\mathrm{TA}, \mathrm{RSSI})$\;
    Remove from $\mathcal{B}$ any entry with $t_{\mathrm{now}} - t_{\mathrm{last}} \geq \tau$\;
    \ForEach{entry $(\boldsymbol{\mu}, c, \tau, t_{\mathrm{last}})$ in $\mathcal{B}$}{
      \If{$d(\mathbf{f}, \boldsymbol{\mu}) \leq \varepsilon$}{
        Reject connection and drop \gls{rrc} context\;
        \textbf{break}\;
      }
    }
  }
}
\end{algorithm}

Overall, the combined detection and mitigation logic blocks new connection attempts that match the fingerprints of suspected \glspl{mue}, while still allowing legitimate \glspl{ue} to access the network. This limits the number of active \gls{rrc} contexts and prevents resource exhaustion. Although attackers may continue to perform connection attempts, these are no longer admitted into the system, and any \gls{rrc} resources previously allocated to malicious requests are automatically released by the \gls{gnb} after the procedure times out. This allows the system to gradually return to a stable state.

\section{Prototype and Experimental Evaluation Setup}
\label{sec:architecture}

To evaluate the performance and robustness of \name in real-world scenarios, we implement its detection and mitigation logic as an O-RAN xApp running on a near-\gls{rt} \gls{ric} based on FlexRIC~\cite{schmidt2021flexric}, and test it into the two \gls{ota} setups shown in Fig.~\ref{fig:hw_setups}.
All software components, including \gls{gnb}, soft-\gls{ue}, \gls{ric}, and xApp, are containerized and orchestrated using Red Hat OpenShift, a Kubernetes-based container platform designed for automated deployment, scaling, and management of applications in hybrid cloud environments. This deployment choice allows us to reuse the same deployment artifacts across both testbed setups and to repeat experiments under controlled resource allocations, as done in~\cite{maxenti2025autoran}.

The first setup, shown in Fig.~\ref{fig:setup_usrp}, leverages an \gls{usrp} X410 \gls{sdr} for signal transmission and reception. The \gls{gnb} protocol stack is implemented using the \gls{oai} framework running within a Red Hat OpenShift pod, with a functional split 8.1.
In this setup, both the \gls{gnb} and the \gls{mue} rely on \gls{sdr} hardware, and we use OctoClock clock distributors to distribute common \gls{pps} and reference signal to the \glspl{sdr} to ensure frame-timing alignment.
This improves the stability of the \gls{oai} softwarized \gls{ue} when connecting to the \gls{sdr}-based \gls{gnb}.

The second testbed setup, illustrated in Fig.~\ref{fig:setup_foxconn}, adopts the 7.2 functional split and integrates a commercial Foxconn \gls{ru}. This is connected to a \gls{gnb} implementation distributed across two OpenShift containers within a single NVIDIA ARC pod. The first container runs the \gls{oai} stack, implementing the \gls{cu} and \gls{du}-high layers, while the second runs NVIDIA cuBB framework for \gls{du}-low functionalities~\cite{villa2025x5g}. 

The NVIDIA ARC pod runs on a Grace Hopper Supermicro ARS-111GL-NHR server with a Grace \gls{cpu} (ARM-based, 72 cores), $512$\,GB of memory, an H100 \gls{gpu}, and a Mellanox ConnectX-7 \gls{nic}, enabling high-throughput networking and hardware acceleration. In contrast, the \gls{oai} \gls{gnb} and \glspl{mue} pods are hosted on a Microway EPYC server equipped with an AMD EPYC 7262 \gls{cpu}, $256$\,GB of \gls{ram}, and a Mellanox ConnectX-6 \gls{nic}.

In both setups, the victim \gls{ue} is a \gls{cots} smartphone (OnePlus Nord) that operates as a standard \gls{5g} device.
The \gls{mue}, instead, is implemented as a software-defined \gls{ue} built on top of the \gls{oai} \gls{ue} protocol stack and driven by a \gls{usrp} X410 device.
We modified the \gls{oai} \gls{ue} \gls{rrc} implementation to repeatedly initiate connection procedures and block them before MSG5, reproducing the signaling storm behavior described in Sec.~\ref{sec:attack}.

The key \gls{gnb} radio parameters used in our setups are summarized in Tab.~\ref{tab:key_radio_params_numeric}.
\vspace{-0.3cm}
\begin{table}[h]
\centering
\small
\setlength{\tabcolsep}{6pt}
\renewcommand{\arraystretch}{1}
\setlength\abovecaptionskip{2pt}{}
\caption{gNB Radio Parameters used for the Experiments}
\begin{tabular}{
>{\raggedright\arraybackslash}l
>{\raggedright\arraybackslash}p{2cm}
>{\raggedright\arraybackslash}p{2cm}
}
\toprule
\textbf{Parameter} & \textbf{USRP X410} & \textbf{Foxconn RU} \\
\midrule
Numerology $\mu$                & 1 & 1 \\
SCS [kHz]          & 30      & 30 \\
PRB                & 162     & 106 \\
Bandwidth [MHz]    & $\sim$60  & $\sim$40 \\
Carrier frequency [GHz] & 3.62928 & 3.75000 \\
Max Power [dBm]
& 23 & 24 \\
\bottomrule
\end{tabular}
\label{tab:key_radio_params_numeric}
\end{table}
\vspace{-0.3cm}
In both setups, the \gls{gnb} connects to a \gls{nearrt} \gls{ric} deployed using FlexRIC, and interfaces with the \name xApp that we developed for detection and mitigation. These are deployed in separate pods exchanging \glspl{kpm} and control decisions with the \gls{gnb} E2 Agent over the O-RAN E2 interface.

\begin{figure}[ht]
  \centering

  \begin{subfigure}[t]{0.265\linewidth}
    \centering
    \includegraphics[width=\linewidth]{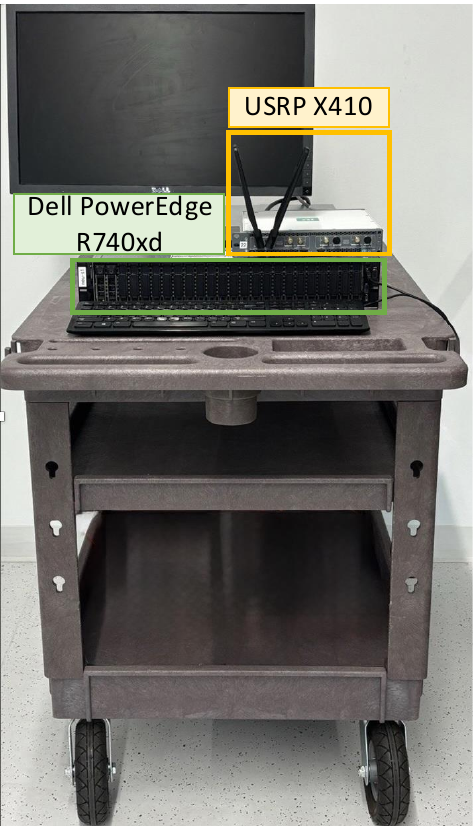}
    \caption{Mobile MUE}
    \label{fig:setup_mobile_mue}
  \end{subfigure}\hfill
  \begin{subfigure}[t]{0.7\linewidth}
    \centering
    \includegraphics[width=\linewidth]{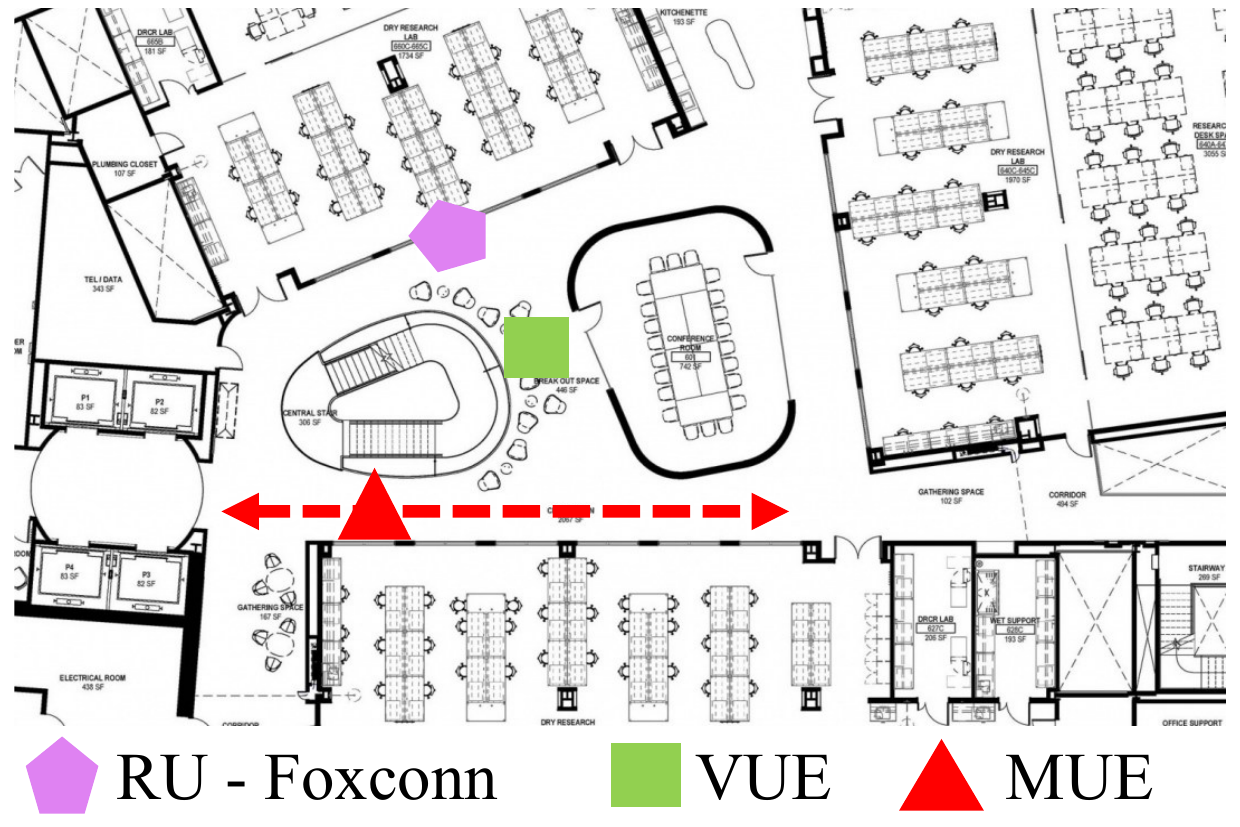}
    \caption{Device Positions during the Experiments}
    \label{fig:floor_map}
  \end{subfigure}

  \vspace{-0.3cm}
  \caption{Mobile MUE Experimental Setup}
  \Description{Mobile MUE experimental setup}
  \label{fig:mobile_mue_combined}
\end{figure}

In addition to these two static setups, we also consider a third configuration used in our mobility experiments. We place an \gls{usrp} X410 and the server driving it on a cart to make the \gls{mue} mobile (Fig.~\ref{fig:setup_mobile_mue}). 
\vspace{-0.2cm}
In this case, the server used to control the \gls{sdr} is a Dell with a 16-core Intel Xeon Gold 6244 \gls{cpu} (hyper-threading disabled), $192$\,GB of \gls{ram}, and a Mellanox ConnectX-6 \gls{nic}. The server runs the modified \gls{oai} \gls{ue} stack that generates the signaling storm behavior while the cart moves, causing the attacker fingerprint $\mathbf{f} = (\mathrm{TA}, \mathrm{RSSI})$ to change over time. \textcolor{black}{Beyond evaluating mobility, this experiment also captures adaptive malicious behaviors that change the observable fingerprint over time (e.g., through transmit power adaptation), since both mobility and power adaptation manifest primarily as \gls{rssi} and, in some cases \gls{ta}, variations determined by wireless propagation conditions.}

Overall, \name follows the O-RAN specification requirements of the O-RAN Use Cases Analysis Report~\cite[Section~4.15]{oran-wg1-use-cases}, which define mechanisms for detecting and mitigating signaling storms caused by misbehaving or \glspl{mue}. These include \gls{nearrt} detection at the \gls{ric} and dynamic mitigation via E2 policies. We implemented the required E2SM and O-RAN procedures by extending both FlexRIC and the \gls{oai} \gls{gnb}, enabling ASN.1 encoded message exchange between the xApp and the \gls{ran}.

\begin{figure}[h]
  \centering
\includegraphics[width=0.69\linewidth]{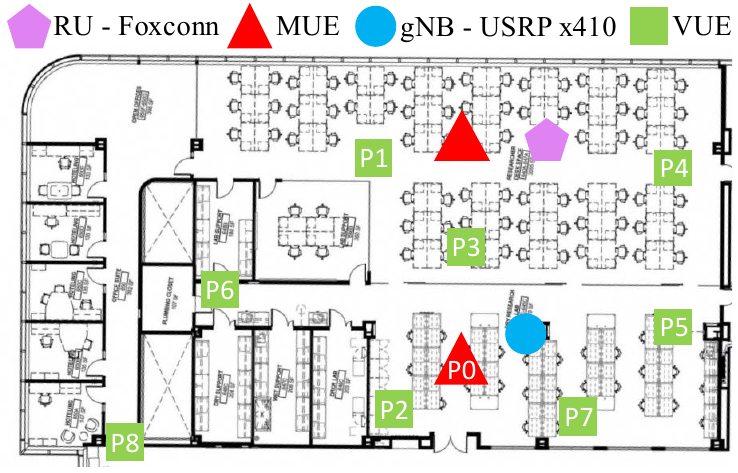}
  \vspace{-0.3cm}
  \caption{Floor Map with Positions of gNBs, MUEs, and VUE}
  \Description{Floor map with locations of gNBs, MUEs, and VUE}
  \label{fig:lab_map}
\end{figure}

Finally, Figs.~\ref{fig:lab_map} and~\ref{fig:floor_map} illustrate the locations of \gls{gnb}, \gls{mue} and \gls{vue} within our indoor testbed for both static and mobile experiments.
In the experiments without mobility (Fig.~\ref{fig:lab_map}), the \gls{vue} is placed at the location denoted with P3. In the mobility experiments (Fig.~\ref{fig:floor_map}), instead, we move the \gls{mue} along the path marked with an arrow in the figure (\textasciitilde $15$\,m) while attacking the \gls{gnb}.
\vspace{-0.2cm}

\section{Experimental Results}
\label{sec:results}

This section presents the experiments conducted to validate the effectiveness of \name detection and mitigation logic. The goal is to demonstrate \name's ability to block \glspl{mue} before they saturate the \gls{gnb} resources, while ensuring that legitimate users can successfully connect to the network.

Table~\ref{tab:parameters} summarizes the key parameters used by the detection and mitigation logic. 
These include the values for the $\MinPts$ parameter that specifies the minimum number of neighbors required to form a valid cluster in the \gls{dbscan} algorithm. We recall that, as discussed in Sec.~\ref{sec:logic}, a neighbor of a point is any other point that lies within a certain distance in the feature space. In our case, each point represents a \gls{ue} fingerprint in the (\gls{rssi}, \gls{ta}) space, and two points are neighbors if their fingerprint distance is below $\epsrssi$ and $\epsta$.
\begin{table}[ht]
\centering
\small
\setlength{\tabcolsep}{6pt}
\renewcommand{\arraystretch}{1}
\setlength\abovecaptionskip{2pt}
\caption{Parameters for Detection and Mitigation Algorithm}
\begin{tabular}{
>{\raggedright\arraybackslash}l
>{\raggedright\arraybackslash}p{4.3cm}
>{\raggedright\arraybackslash}l
}
\toprule
\textbf{Parameter} & \textbf{Description} & \textbf{Value} \\
\midrule
$\epsrssi$   & Max RSSI difference                  & 4 \\
$\epsta$     & Max Timing Advance difference        & 1 \\
$\MinPts$    & Min. points for \gls{dbscan} cluster & 3 \\
$T_1$        & MSG3 count threshold                 & 3 \\
$T_2$        & Lower limit for $R_1/R_2$ metrics    & 0.25 \\
$T_3$        & Cluster density threshold            & Dynamic \\
$W$          & Time window for detection            & $100$\,ms \\
\glspl{mue}  & Number of attacker UEs               & 1--2 \\
Victim UEs   & Number of victim UEs                 & 1 \\
\bottomrule
\end{tabular}
\label{tab:parameters}
\end{table}
\textcolor{black}{We emphasize that this fingerprint is not intended for unique device identification or fine-grained localization: \gls{ta} provides a coarse proximity estimation, while \gls{rssi} captures channel and hardware-dependent effects (e.g., LoS/NLoS and interference). Both $\epsrssi$ and $\epsta$ are set from the maximum short-term \gls{rssi} and \gls{ta} variation observed for a static \gls{ue}.}
$T1$ defines the maximum number of MSG3 messages allowed within a time window. Exceeding this threshold suggests potential malicious activity, and if the $T2$ and $T3$ thresholds are also exceeded, the system confirms an attack.
We test our approach in the static and mobile configurations described in Sec.~\ref{sec:architecture}, with both single and multiple attackers, and also evaluate benign \gls{ue} connecting from different locations (as described in Fig.~\ref{fig:lab_map}). \textcolor{black}{To evaluate collateral blocking cases, each run includes both the attacking \glspl{mue} and a legitimate victim \gls{ue} attempting to connect to the \gls{gnb} under attack. Additionally, we conduct benign-only experiments to quantify the TN and FP, using the same number of runs as in the attack scenarios.}

\begin{figure}[ht]
  \centering  \includegraphics[width=1\linewidth, height=0.25\textheight]{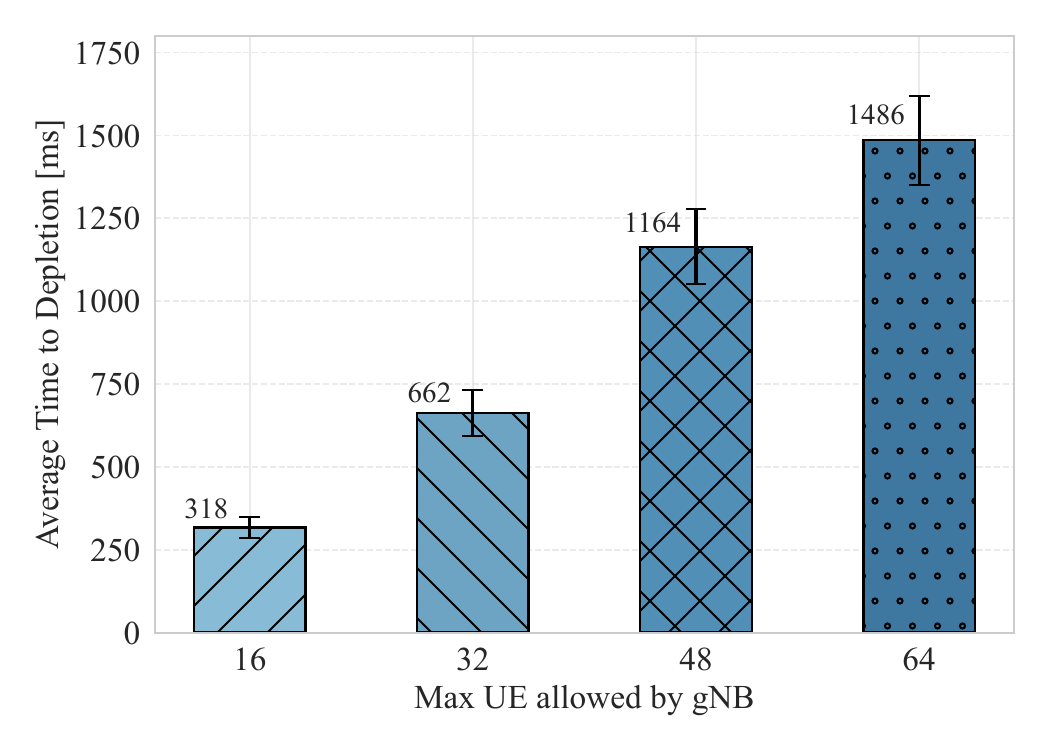}
  \vspace{-1cm}
  \caption{Resource Depletion Time vs.\ Max. \#UEs Connected to the gNB}
  \Description{Resource depletion time vs.\ max. \#UEs at gNB}
  \label{fig:barplot_time_vs_maxUE}
\vspace{-0.5cm}
\end{figure}

To quantify the impact of an \gls{rrc} signaling storm, we measure the time-to-depletion of \gls{rrc} contexts as a function of the \gls{gnb}’s configured Max~UE (the upper bound on simultaneous \gls{rrc} contexts). The detection/mitigation xApp is disabled, and one \gls{mue} issues \gls{rrc} Setup Requests at an average rate of 45.7~MSG3/s. By default, \gls{oai} sets Max~UE to 16, but we vary it in \{16, 32, 48, 64\} in our experiments.
This result is shown in Fig.~\ref{fig:barplot_time_vs_maxUE}, where each bar is averaged over ten experiment runs. We notice that the time-to-depletion grows linearly with the Max~UE parameter.
This shows that simply configuring the \gls{gnb} to allow more concurrent \gls{ue} connections only postpones exhaustion but it does not prevent it. Indeed, the \gls{gnb} remains vulnerable to the attack, which demonstrates the need for our designed \name detection and mitigation techniques to proactively identify and block malicious access.

\begin{table*}[!t]
\color{black}
\centering
\small
\setlength{\tabcolsep}{4pt}
\renewcommand{\arraystretch}{1}
\setlength\abovecaptionskip{2pt}
\caption{Detection and Mitigation Evaluation Across Different Scenarios}
\begin{tabularx}{\textwidth}{
  >{\raggedright\arraybackslash}p{4.5cm}
  >{\centering\arraybackslash}p{2cm}
  >{\centering\arraybackslash}p{2cm}
  *{5}{>{\centering\arraybackslash}p{1cm}}
  >{\centering\arraybackslash}p{1.5cm}
}
\toprule
\textbf{Scenario} & \textbf{Radio} & \textbf{\#\,Experiments} &
\textbf{TP} & \textbf{FP} & \textbf{TN} & \textbf{FN} &
\textbf{Accuracy} & \textbf{CB} \\
\midrule
1 Static VUE \& 1 Static MUE & USRP       & 12 & 100 \% & 0 \% & 100 \% & 0 \% & 100 \% & 25 \% \\
1 Static VUE \& 2 Static MUE & USRP       & 12 & 92 \% & 0 \% & 100 \% & 8 \% & 96 \% & 17 \% \\
1 Static VUE \& 1 Static MUE & Foxconn RU & 12 & 100 \% & 0 \% & 100 \% & 0 \% & 100 \% & 17 \% \\
1 Static VUE \& 2 Static MUE & Foxconn RU & 12 & 92 \% & 0 \% & 100 \% & 8 \% & 96 \% & 17 \% \\
1 Static VUE \& 1 Mobile MUE & Foxconn RU & 12 & 92 \% & 0 \% & 100 \% & 8 \% & 96 \% & 25 \% \\

\midrule
\multicolumn{3}{c}{} &
\multicolumn{6}{c}{\textbf{Success Rate} } \\
\cmidrule(lr){4-9}
\multicolumn{3}{c}{} &
\textbf{P1} & \textbf{P2} & \textbf{P3} & \textbf{P4} & \textbf{P5} & \textbf{Average} \\
1 Mobile VUE \& 2 Static MUEs & USRP       & 12 per location & 60 \% & 100 \% & 80 \% & 90 \% & 70 \% & 80 \% \\
1 Mobile VUE \& 2 Static MUEs & Foxconn RU & 12 per location & 70 \% & 80 \% & 80 \% & 90 \% & 70 \% & 80 \%\\

\bottomrule
\end{tabularx}
\label{tab:summary_results}
\end{table*}

Next, we enable the detection and mitigation logic of \name and test three scenarios: (i) static attacker and static victim, (ii) static attacker and mobile victim, and (iii) mobile attacker and static victim. Results are shown in Tab.~\ref{tab:summary_results}, where for each case, we collect detection statistics over 12 experiment runs.
The classification outcomes are defined as:
\textcolor{black}{
\begin{itemize}
\item \textbf{True Positive (TP):} an attack is correctly detected and blocked before resource depletion.
\item \textbf{True Negative (TN):} no attack is present and the system correctly remains inactive, allowing benign \glspl{ue} to connect.
\item \textbf{False Positive (FP):} no attack is present, but the system mistakenly triggers mitigation.
\item \textbf{False Negative (FN):} an attack is present but not detected or mitigated, and \glspl{mue} deplete the \gls{gnb} resources.
\end{itemize}
}
\textcolor{black}{In addition, we define the \textbf{\gls{cbr}} as the percentage of benign \glspl{ue} that are temporarily impacted while an ongoing attack is correctly detected and mitigated, e.g., because of fingerprint similarity with the \glspl{mue}.}

\begin{table}[H]
\centering
\color{black}
\small
\setlength{\tabcolsep}{4pt}
\renewcommand{\arraystretch}{1}
\setlength\abovecaptionskip{2pt}
\caption{\textcolor{black}{Fine Tuning of Detection and Mitigation Parameters}}
\label{tab:fine_tuning}

\begin{tabularx}{\columnwidth}{
  >{\centering\arraybackslash}p{0.4cm}   
  >{\centering\arraybackslash}p{0.5cm}   
  >{\centering\arraybackslash}p{0.4cm}   
  >{\centering\arraybackslash}p{0.4cm}   
  >{\centering\arraybackslash}p{0.4cm}   
  >{\centering\arraybackslash}p{0.7cm}   
  >{\centering\arraybackslash}p{0.7cm}   
  >{\centering\arraybackslash}p{0.7cm}   
  >{\centering\arraybackslash}p{0.7cm}   
  >{\centering\arraybackslash}p{0.7cm}   
}
\toprule
\textbf{Exp.} &
$\boldsymbol{\epsilon_{\mathrm{RSSI}}}$ &
$\boldsymbol{\epsilon_{\mathrm{TA}}}$ &
$\boldsymbol{T_1}$ &
$\boldsymbol{T_2}$ &
\textbf{TP} &
\textbf{FP} &
\textbf{TN} &
\textbf{FN} &
\textbf{CB} \\
\midrule
E1 & 4  & 1 & 3  & 0.25 & 100\% & 0\%   & 100\% & 0\%  & 0\% \\
E2 & 1  & 1 & 3  & 0.25 & 86\%  & 0\%   & 100\% & 14\% & 0\% \\
E3 & 10 & 1 & 3  & 0.25 & 86\%  & 0\%   & 100\% & 14\% & 71\% \\
E4 & 4  & 1 & 1  & 0.25 & 100\% & 14\%  & 86\%  & 0\%  & 0\% \\
E5 & 4  & 1 & 10 & 0.25 & 29\%  & 0\%   & 100\% & 71\% & / \\
E6 & 4  & 1 & 1  & 1    & 100\% & 100\% & 0\%   & 0\%  & 0\% \\
E7 & 4  & 1 & 10 & 1    & 29\%  & 0\%   & 100\% & 71\% & / \\
\bottomrule
\end{tabularx}
\end{table}
\textcolor{black}{
In order to identify a robust evaluation setting for detection and mitigation, we fine-tune the fingerprint clustering parameters
($\epsilon_{\text{RSSI}}$, $\epsilon_{\text{TA}}$) and the attack detection parameters ($T_1$, $T_2$), and evaluate the resulting trade-offs in TP/TN/FP/FN and \gls{cbr} as shown in Table~\ref{tab:fine_tuning}.
Configuration \textbf{E1} provides the best configuration: it blocks the \gls{mue} before resource saturation (TP=100\%, FN=0\%) while
preserving benign traffic (FP=0\%) and avoiding collateral blocking (CB=0\%).
\textbf{E2} appears competitive, but the very small $\epsilon_{\text{RSSI}}$ makes clustering too sensitive, so minor attacker fingerprint variations can fragment the malicious fingerprints into multiple clusters.
\textbf{E3} shows the opposite effect: a large $\epsilon_{\text{RSSI}}$ may merge the victim fingerprint with the malicious cluster, leading to high collateral blocking even without direct misclassification (FP=0\%, CB=71\%).
\textbf{E4} achieves high detection (TP=100\%) but all benign traffic is classified as high-load, even with a single \gls{vue}.
\textbf{E6} is overly aggressive due to a too-low $T_1$, resulting in a high FP rate (FP=100\%).
Finally, \textbf{E5} and \textbf{E7} are too conservative (high $T_1$), missing most attacks (FN=71\%); in these cases, \gls{cbr} is reported as / because the attacker saturates all resources, preventing the \gls{vue} from connecting and thus making collateral blocking impossible to measure.}
\textcolor{black}{In light of the above considerations, we evaluate \name across the scenarios of Table~\ref{tab:summary_results} using the best-performing configuration identified during fine-tuning phase, i.e., E1.}
We notice that in the case of a single \gls{mue} and \gls{usrp} \gls{gnb}, \textcolor{black}{\name reaches a detection accuracy of 100\% with a 25\% \gls{cbr}}. With two \glspl{mue}, instead, the attackers are blocked in the 92\% of cases, with 8\% false negatives (i.e., a single false negative in our case). \textcolor{black}{In 17\% of the runs, the benign \gls{ue} was unable to connect during an attack that was correctly blocked by our system}.
The setup with the Foxconn \gls{gnb} presents a similar behavior. With a single \gls{mue}, attacks are mitigated in all runs. With two \glspl{mue}, attacks are blocked in the 92\% of runs, with 8\% of missed detections. \textcolor{black}{The benign \glspl{ue} are collaterally blocked in 17\% of the cases for both the scenarios.}
This consistency across the \gls{usrp} and Foxconn \gls{ru} setups is expected. The attack and our detection logic operate at the \gls{rrc} control-plane layer, which is the same in both stacks, while the two testbeds mainly differ in the lower \gls{phy}/\gls{rf} implementation (i.e., \gls{sdr} vs.\ commercial \gls{ru} and split~8.1 vs.\ 7.2). As a result, hardware and split choices introduce only minor quantitative differences but do not change the qualitative behavior of the attack or of the proposed solution. Hence, we report per-radio results to highlight robustness across heterogeneous environments.

\begin{figure}[ht]
  \centering
  \includegraphics[width=1\linewidth]{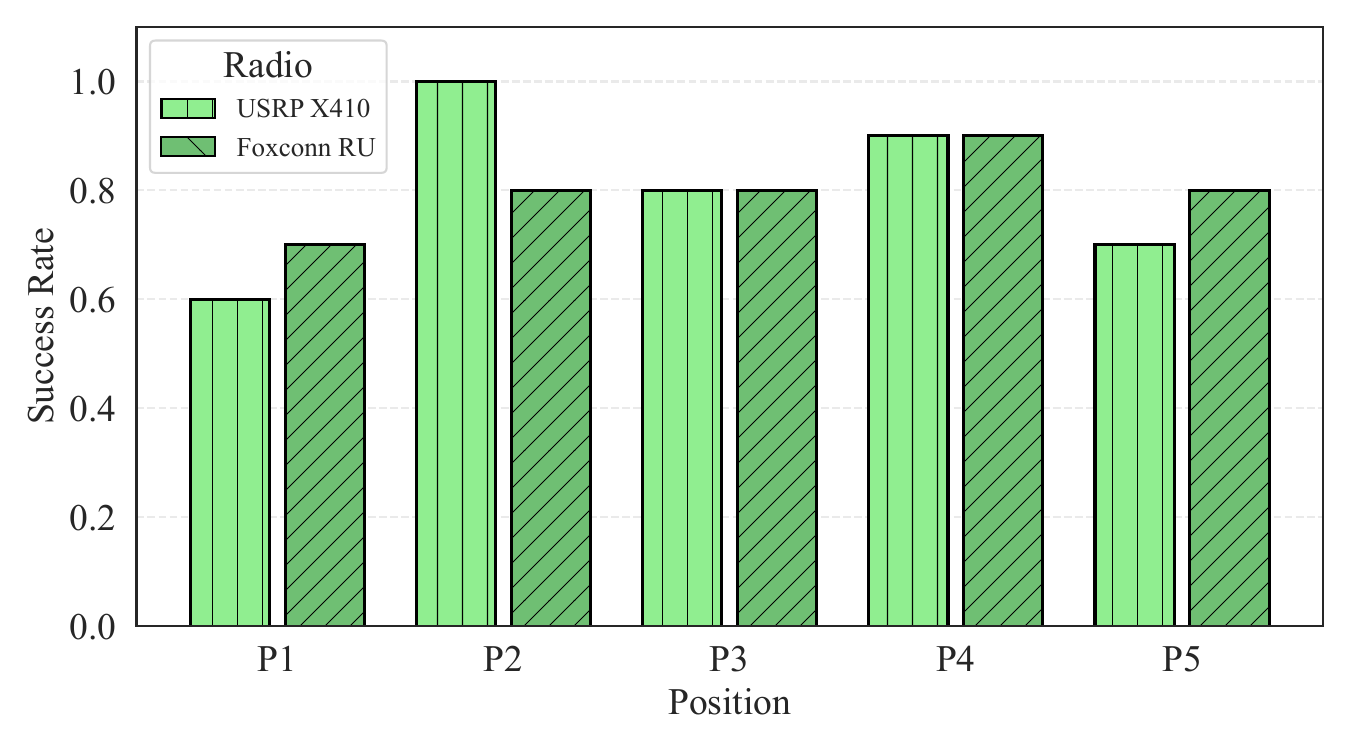}
  \vspace{-0.8cm}
  \caption{\textcolor{black}{Success Rate Across Positions P1-P5}}
  \Description{Success Rate Across Locations P1-P5}
  \label{fig:accuracy_comparison_per_position}
\end{figure}

Results for the mobile experiments demonstrate our logic blocks the mobile attacker in 92\% of the runs, missing one case (i.e., 8\%), \textcolor{black}{while the benign \gls{ue} is not allowed to connect during a mitigated attack in the 25\% of cases}. 
These values confirm that \name can effectively handle mobile threats, \textcolor{black}{though mobility slightly increases the \gls{cbr}} compared to static scenarios.
We then evaluate the system with a mobile \gls{vue} attempting to connect from locations P1–P5 of Fig.~\ref{fig:lab_map}, \textcolor{black}{while two static \glspl{mue} continuously perform a signaling storm attack against the \gls{gnb}}.
The goal of this experiment is to assess spatial variability in detection accuracy during mitigated attacks.
Figure~\ref{fig:accuracy_comparison_per_position} shows the \textcolor{black}{success rate} (first-attempt access without being blocked by the detection and mitigation system) at each location represented as green squares in Fig.~\ref{fig:lab_map}.
We notice that the success rate remains consistently high across all locations, though slight variations are observed. Specifically, \textcolor{black}{locations P1, P3, and P5 show higher values for \gls{cbr}, with success rate dropping below 80\%, while P2 and P4 maintain success rate above 80\%. This behavior can be explained by the fact that fingerprint similarity is not determined only by the spatial proximity, but by the joint $(\mathrm{TA},\mathrm{RSSI})$ signature: the \gls{ta} primarily captures the \gls{ue} radial distance from the gNB, whereas the \gls{rssi} is also affected by propagation conditions and interference. As a result, some victim locations (P1, P3, and P5) experience radio conditions yielding \gls{rssi} values closer to those of the \glspl{mue}, thus increasing fingerprint similarity and the likelihood of misclassification. Conversely, although P2 appears close to the attacker in Fig.~\ref{fig:lab_map}, its \gls{rssi} remains sufficiently different, resulting in a lower \gls{cbr}.}
\vspace{-0.3cm}
\vspace{-0.3cm}
We summarize the results of the previous experiments in Tab.~\ref{tab:summary_results} to provide a comprehensive overview of \name capabilities.

\textcolor{black}{To evaluate fingerprint robustness and \gls{rssi} stability, we conduct additional experiments to (i) assess the separability of the fingerprint across multiple locations and (ii) quantify its stability under fixed transmission conditions. 
We place the \gls{vue} at 8 distinct fixed positions (VUE--P1 to VUE--P8) and collect 10 experiment runs per position to compute mean and standard deviation; we also test a malicious soft-UE on a \gls{usrp}~X410 (MUE--P0) to assess the impact of hardware heterogeneity. 
\vspace{-0.1cm}
\begin{table}[ht]
\color{black}
\centering
\small
\setlength{\tabcolsep}{6pt}
\renewcommand{\arraystretch}{1}
\setlength\abovecaptionskip{2pt}{}
\caption{\textcolor{black}{Fingerprint for an UE at different Fixed Positions}}
\begin{tabular}{
  >{\raggedright\arraybackslash}l
  >{\raggedright\arraybackslash}p{2cm}
  >{\raggedright\arraybackslash}p{2cm}
}
\toprule
\textbf{Position} & \textbf{TA [samples]} & \textbf{RSSI [dBm]} \\
\midrule
MUE - P0 & $32.0 \pm 0.0 $ & $-41.0 \pm 0.0$ \\
VUE - P1 & $31.1 \pm 0.5 $ & $-49.7 \pm 3.2$ \\
VUE - P2 & $31.0 \pm 0.0 $ & $-56.4 \pm 4.4$ \\
VUE - P3 & $31.0 \pm 0.0 $ & $-52.6 \pm 1.2$ \\
VUE - P4 & $30.9 \pm 0.3 $ & $-53.6 \pm 1.8$ \\
VUE - P5 & $30.9 \pm 0.3 $ & $-54.8 \pm 1.0$ \\
VUE - P6 & $31.0 \pm 0.0 $ & $-46.7 \pm 3.7$ \\
VUE - P7 & $31.2 \pm 0.4 $ & $-55.0 \pm 0.0$ \\
VUE - P8 & $31.1 \pm 0.5 $ & $-47.5 \pm 1.3$ \\
\bottomrule
\end{tabular}
\label{tab:fp_stability}
\end{table}
Table~\ref{tab:fp_stability} reports the resulting statistics. We observe that the mean \gls{rssi} varies enough to distinguish the 8 different victim locations (median inter-location difference $\approx 3.9$\,dB, up to 9.7\,dB), while remaining stable within each location (median $\sigma \approx 1.5$\,dB). Finally, the \gls{usrp}-based attacker exhibits a very different fingerprint with better \gls{rssi}. This difference is expected because the attacker uses an \gls{sdr} platform with transmission characteristics that differ from \gls{cots} phones and, in our setup, it is placed very close to the \gls{gnb} under line-of-sight conditions, resulting in lower path loss and thus higher received power.
This suggests that hardware heterogeneity and different propagation conditions in larger \gls{5g} deployments can further improve fingerprint separability, even in restricted environments.}

\begin{figure}[ht]
  \centering
  \includegraphics[width=1\linewidth]{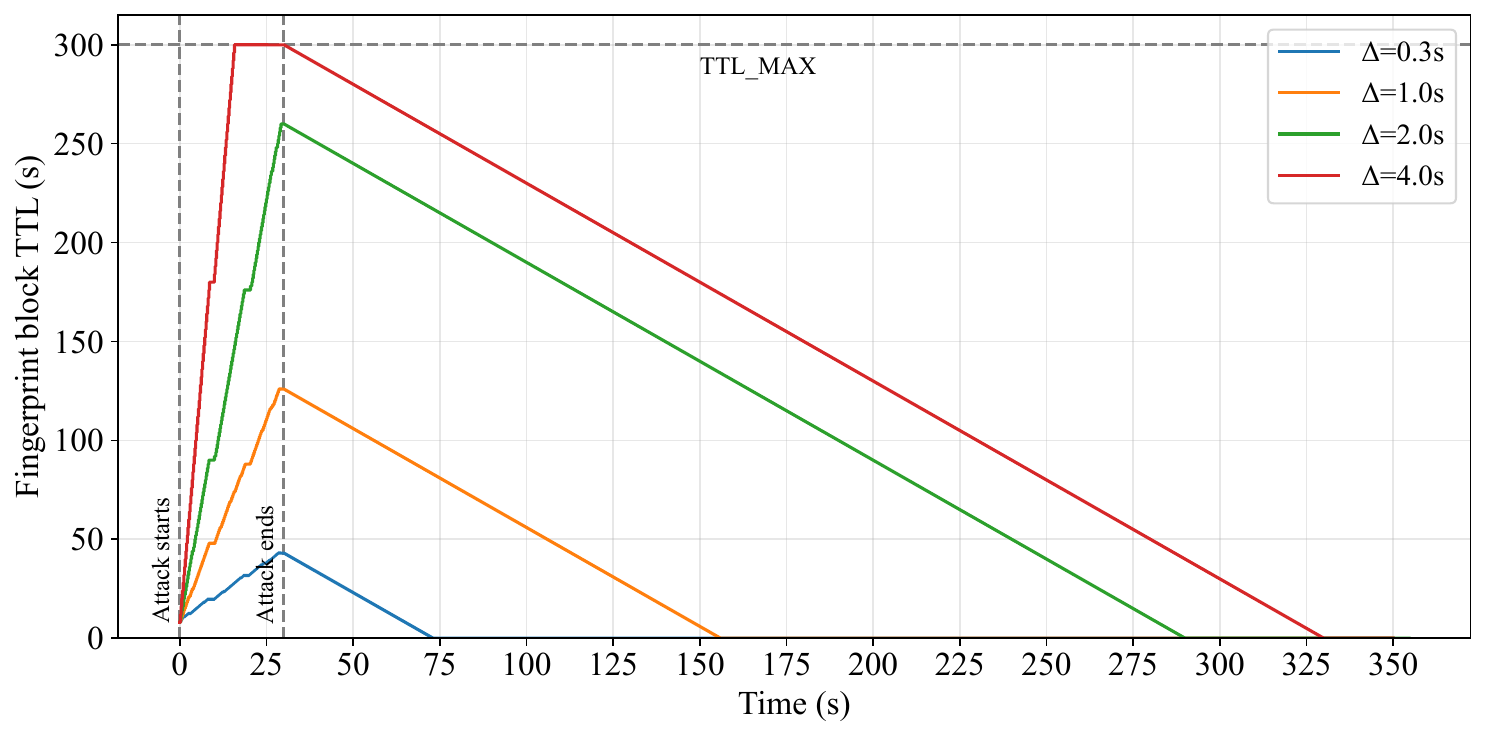}
  \vspace{-0.8cm}
  \caption{\textcolor{black}{Aging Timeout Evolution vs. Windows $\Delta$}}
  \Description{Aging timeout evolution vs. windows $\Delta$}
  \label{fig:aging_mechanism_real}
\end{figure}

\textcolor{black}{In addition, we analyze the proposed fingerprint aging mechanism, as its timeout directly impacts post-attack residual blocking and recovery time.
Figure~\ref{fig:aging_mechanism_real} reports the evolution of the mitigation timeout for a single malicious fingerprint during a signaling storm attack of fixed duration ($30$\,s) using real traces collected at the \gls{gnb}. Each curve corresponds to a different aging window (increment step) $\Delta$: larger $\Delta$ values make the timeout grow faster, producing a longer residual blocking period after the attack ends, whereas smaller $\Delta$ values yield a slower growth and therefore a faster recovery. Specifically, with the largest $\Delta$, the timeout caps at $TTL_{\max}$ within the $30$\,s attack window and then remains constant at $TTL_{\max}$ for the remainder of the attack. Conversely, for smaller $\Delta$ settings, the timeout does not reach the cap within $30$\,s and the curves keep increasing without saturating. After the attack ends (no further detector refreshes), all curves go to zero, with the time-to-zero directly quantifying the recovery time. Indeed, configurations that saturate at (or approach) $TTL_{\max}$ enforce a more conservative mitigation and longer post-attack blocking, while smaller $\Delta$ provide a more forgiving behavior with shorter expiration periods and faster restoration of service availability for the blocked fingerprints.}

\vspace{-0.2cm}
\begin{figure}[ht]
  \centering
  \includegraphics[width=1\linewidth]{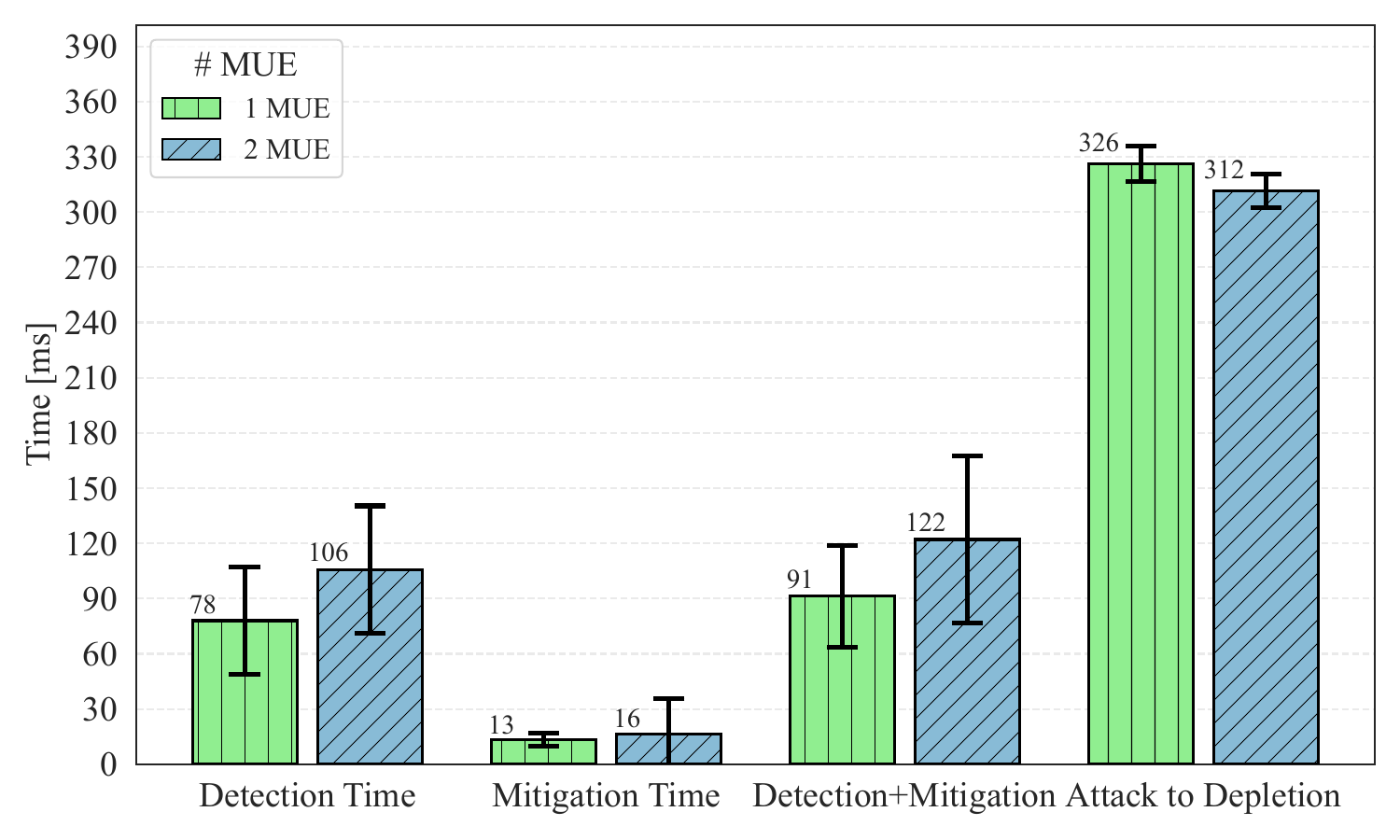}
  \vspace{-0.8cm}
  \caption{Detection and Mitigation vs Depletion Times}
  \Description{Detection and mitigation vs depletion times}
  \label{fig:usrp_latency_comparison__multiple_mue}
\end{figure}

\vspace{-0.2cm}
Finally, we measure different time metrics to assess the performance of the detection and mitigation process: the \textbf{detection time}, i.e., the time required by the xApp to recognize an ongoing attack and detect the presence of a \gls{mue}; the \textbf{mitigation time}, namely the time needed to apply countermeasures once detection occurs; and the \textbf{depletion time}, defined as the time until one or more \glspl{mue} exhaust the \gls{gnb} resources in the absence of the detection/mitigation xApp.
Figure~\ref{fig:usrp_latency_comparison__multiple_mue} shows the results of two sets of experiments with one and two \glspl{mue}. Each bar corresponds to the average over 12 experiment runs.
With one static attacker, detection and mitigation occur in $91$\,ms, on average, against a $326$\,ms depletion time. 
With two \glspl{mue}, instead, detection and mitigation happen in $122$\,ms, which is still below the $312$\,ms depletion latency.
This shows that detection and mitigation times are significantly lower than depletion time, demonstrating that \name is fast and effective in preventing \gls{gnb} resource exhaustion.
\vspace{-0.3cm}
\vspace{-0.3cm}
The results presented in this section confirm the robustness of the proposed \name system that can consistently block \glspl{mue} before exhausting \gls{gnb} resources, with only minor degradation in accuracy in the case of mobility.

\section{Conclusions}
\label{sec:conclusions}

In this work, we considered the problem of \gls{rrc} Signaling Storm attacks in \gls{5g} networks, which can rapidly exhaust \gls{gnb} resources by flooding connection requests in the early stages of the \gls{rrc} connection procedure.
Through an experimental campaign on two \gls{ota} testbed setups, we first demonstrated that \gls{rrc} signaling storm attacks can be reproduced and studied on realistic \gls{5g} systems with single/multiple, static and mobile attackers.
Then, we proposed \name, a fingerprint-based detection and mitigation approach that combines \gls{rrc} statistics with $(\mathrm{TA},\mathrm{RSSI})$ fingerprints to distinguish attacks from high-load network conditions, and block malicious access attempts before \gls{gnb} resources are exhausted.
We implemented \name as an xApp running on an O-RAN \gls{nearrt} \gls{ric} for closed-loop \gls{gnb} control through a \gls{sm} implemented as described in an O-RAN use case.

By running $\approx$ \textcolor{black}{389 experiments} on a real-world testbed, we demonstrated that our system achieved an average accuracy of \textcolor{black}{97.6\%} in identifying and blocking \glspl{mue}, including in scenarios with mobility and multiple attackers. Detection and mitigation consistently occurred before \gls{gnb} resource depletion, with an average response time of $106.5$\,ms, which shows the effectiveness of \name.

As future work, we plan to improve the accuracy of \gls{ue} fingerprinting by integrating additional features, such as I/Q sample analysis to capture hardware-specific \gls{rf} impairments and provide a more robust device signature. We also aim to expand the detection surface from a local-cell approach to a distributed system leveraging the centralized abstraction of the \gls{ric}. In addition, we will validate the proposed fingerprint in outdoor deployments, where \gls{ta} granularity better separates spatial regions. Finally, we will extend the threat model to include a larger population of benign victim \glspl{ue} to capture high-load scenarios, as well as more advanced, adaptive, and stealthy adversaries capable of completing \gls{rrc} and \gls{nas} authentication and dynamically modifying transmit power to evade detection.

\begin{acks}
Noemi Giustini carried out this work while visiting Northeastern University as a Master's student from Sapienza University of Rome.
This article is based upon work partially supported by the U.S. National Science Foundation under award CNS-2434081 and by OUSD(R\&E) through Army Research Laboratory Cooperative Agreement Number W911NF-24-2-0065. The views and conclusions contained in this document are those of the authors and should not be interpreted as representing the official policies, either expressed or implied, of the Army Research Laboratory or the U.S. Government. The U.S. Government is authorized to reproduce and distribute reprints for Government purposes notwithstanding any copyright notation herein.
The work was also partially supported by SERICS (PE00000014) 5GSec project, CUP B53C22003990006, under the MUR National Recovery and Resilience Plan funded by the European Union - NextGenerationEU.
\end{acks}

\balance
\bibliographystyle{ACM-Reference-Format}

\bibliography{biblio}

\end{document}